\documentclass[12pt]{iopart}
\usepackage{iopams}

\usepackage[english]{babel}
\usepackage[latin2]{inputenc}
\usepackage{t1enc}
\usepackage{graphicx}

\usepackage[normalem]{ulem}
\usepackage{color}
\definecolor{blue}{rgb}{0,0,1}
\definecolor{red}{rgb}{1,0,0}

\newcommand{\R}{\mathbb{R}}
\newcommand{\F}{\mathcal{F}}
\newcommand{\g}{\mathsf{g}}
\newcommand{\N}{\mathcal{N}}

\begin{document}

\title[GR effects in muon g-2, EDM and other spin precession experiments]{Quantification of GR effects in muon g-2, EDM and other spin precession experiments}

\author{Andr\'as L\'aszl\'o}
\address{Wigner Research Centre for Physics, Budapest}
\ead{laszlo.andras@wigner.mta.hu}

\author{Zolt\'an Zimbor\'as}
\address{Wigner Research Centre for Physics, Budapest}
\ead{zimboras.zoltan@wigner.mta.hu}

\begin{abstract}
Recently, Morishima, Futamase and Shimizu published a series 
of manuscripts, putting forward arguments, based on a post-Newtonian 
approximative calculation, that there can be a sizable general 
relativistic (GR) correction in the experimental determination of the muon 
magnetic moment based on spin precession, i.e., in muon g-2 experiments. In response, 
other authors argued that the effect must be much smaller than 
claimed. Further authors argued that the effect exactly cancels. 
Also, the known formulae for de Sitter and Lense-Thirring effect do not 
apply due to the non-geodesic motion.
All this indicates that it is difficult to estimate from first principles 
the influence of GR corrections in the problem of spin propagation. 
Therefore, in this paper we present a full general relativistic calculation 
in order to quantify this effect. The main methodology is the purely 
differential geometrical tool of Fermi-Walker transport over a Schwarzschild 
background. Also the Larmor precession due to the propagation in the 
electromagnetic field of the experimental apparatus is included. 
For the muon g-2 experiments the GR correction turns out to be very small, 
well below the present sensitivity. However, 
in other similar storage ring experimental settings, such as electric dipole 
moment (EDM) search experiments, where the so-called frozen spin method is used, 
GR gives a well detectable effect, and should be corrected for. All frozen 
spin scenarios are affected which intend to reach a sensitivity of 
0.1 microradians/second for the spin precession in the 
vertical plane.
\end{abstract}


\maketitle

\section{Introduction}
\label{secIntroduction}

In a recent series of papers \cite{morishima2018a, morishima2018b, morishima2018c},
it was claimed that, in the muon anomalous 
magnetic moment experiments \cite{g2, bennett2006, miller2007, mane2005}, there can be a general relativistic (GR) correction to the  precession effect of the muon spin direction vector 
when orbiting in the magnetic storage ring sitting on the Earth's surface in a
Schwarzschild metric. These calculations were based on a post-Newtonian 
approximation, and the authors claimed that the pertinent effect may cause 
an unaccounted systematic error in the measurement of the muon's anomalous 
magnetic moment, often referred to as g-2. 
Other papers \cite{visser2018, guzowski2018} responded that the effect 
is much smaller. Further papers \cite{nikolic2018} responded that the effect 
exactly cancels. Moreover, the usual formulae of de Sitter and Lense-Thirring 
precession \cite{lammerzahl2001} do not apply, since the pertinent orbit is non-geodesic. 
All this suggests that it is relatively difficult to say 
something from first principles on the magnitude of GR corrections for 
spin transport in a gyroscopic motion along a forced orbit. 
Motivated by these, in the present paper, we intend to quantify 
the pertinent effect in the context of GR. 
We use the differential geometrical tool 
of Fermi-Walker transport of vectors along trajectories in spacetime. 
In this way, the kinematic precession, called the Thomas precession, can be 
quantified over the Schwarzschild background field of the Earth. This is then 
compared to the Minkowski limit, i.e., when GR is neglected. The effect of 
the Larmor precession in the electromagnetic field of the experimental 
setting is also quantified, and its corresponding GR correction is also 
evaluated. The calculations show that the GR corrections for 
the actual g-2 experimental setting \cite{g2, bennett2006, miller2007, mane2005} 
is very small, well below the experimental sensitivity. There are, however, 
other spin precession experiments, such as the electric dipole moment (EDM) 
search experiments \cite{senichev2017, semertzidis2016, talman2017}, where it turns out that GR gives a rather large signal. Since these 
experiments are intended as sensitive probes for Beyond Standard Model (BSM) 
scenarios, their experimental data should be corrected for the GR effect. 
In particular, the EDM experiments \cite{senichev2017, semertzidis2016, talman2017} could be thought of also as 
sensitive GR experiments on spin propagation of elementary particles, 
kind of microscopic versions of Gravity Probe B \cite{grav, everitt2011} gyroscope experiment. 
During the past years there have been a few papers warning about the possibility of such an 
effect \cite{kobach2016, silenko2007, obukov2016a, obukov2016b, orlov2012}. These estimations, 
however, are not fully covariant Lorentz geometric GR calculations, but are mostly 
special relativistic or semi-general relativistic, or applying other kind of 
approximations such as not fully taking into account GR for the 
electrodynamic part. 
As a result, the estimations 
\cite{kobach2016, silenko2007, obukov2016a, obukov2016b} differ from our geometric GR 
calculation in the details of the particle velocity dependence, and with some 
factors. The post-Newtonian estimation of \cite{orlov2012}, where the GR effect 
for the special case of a purely electric frozen spin storage ring is 
quantified, is confirmed by our covariant calculations. 
Since the EDM signal is expected to sit on this large GR 
background of the order of $30\,\mathrm{nrad/sec}$, the pertinent factors 
matter a lot for discrimination from a BSM signature with the planned 
precision of $1\,\mathrm{nrad/sec}$. The GR signal, however, can also be disentangled 
from a true EDM signal due to their opposite space reflection behavior, i.e., 
by switching beam direction.

The structure of the paper is as follows. Section~\ref{secSetting} outlines the kinematic 
setting of our model of the experimental situation. In Section~\ref{secFWtransport} and \ref{secRelFWtransport}, we discuss, 
from a geometrical point of view, the Fermi-Walker transport (gyroscopic equation) and the general relativistic Thomas precession, respectively.
The GR corrections to the Thomas 
precession is evaluated in Section~\ref{secEvaluationT}. 
In Section~\ref{secEldin}, the idealized model of electromagnetic fields 
in an electromagnetic storage ring over a Schwarzschild background is outlined, 
and their Fermi-Walker-Larmor spin transport is evaluated in 
Section~\ref{secAbsLarmor}. The analytical formulae are derived for a combined 
Thomas and Larmor precession over the Schwarzschild background in Section~\ref{secRelLarmor}.
Finally, in Section~\ref{secEvaluationLarmor} the total GR corrections are evaluated, which is  followed by our concluding remarks in
Section~\ref{secConclusion}.

\section{The kinematic setting}
\label{secSetting}

\begin{figure}[!t]
\begin{center}
\includegraphics[width=3.5cm]{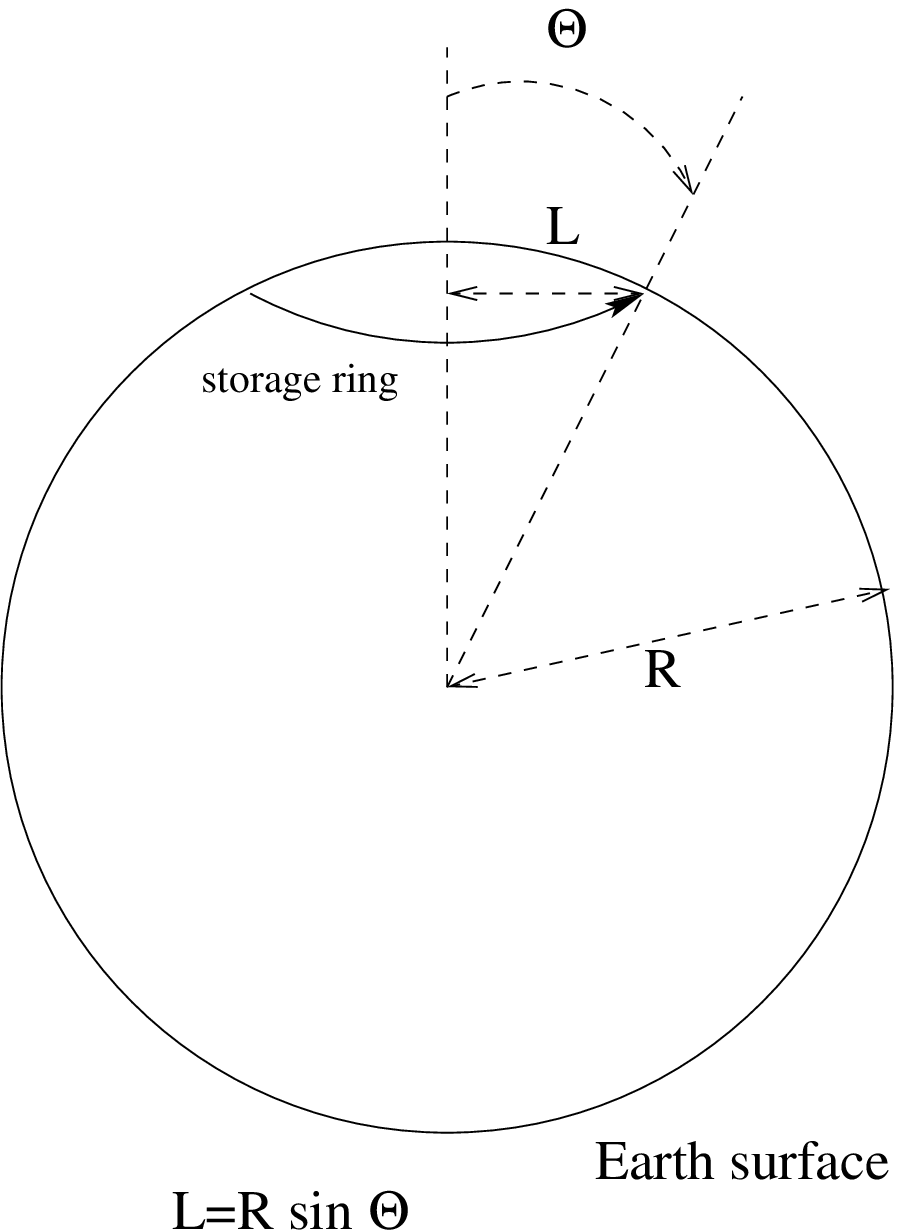}

\vspace*{1cm}
\includegraphics[width=7cm]{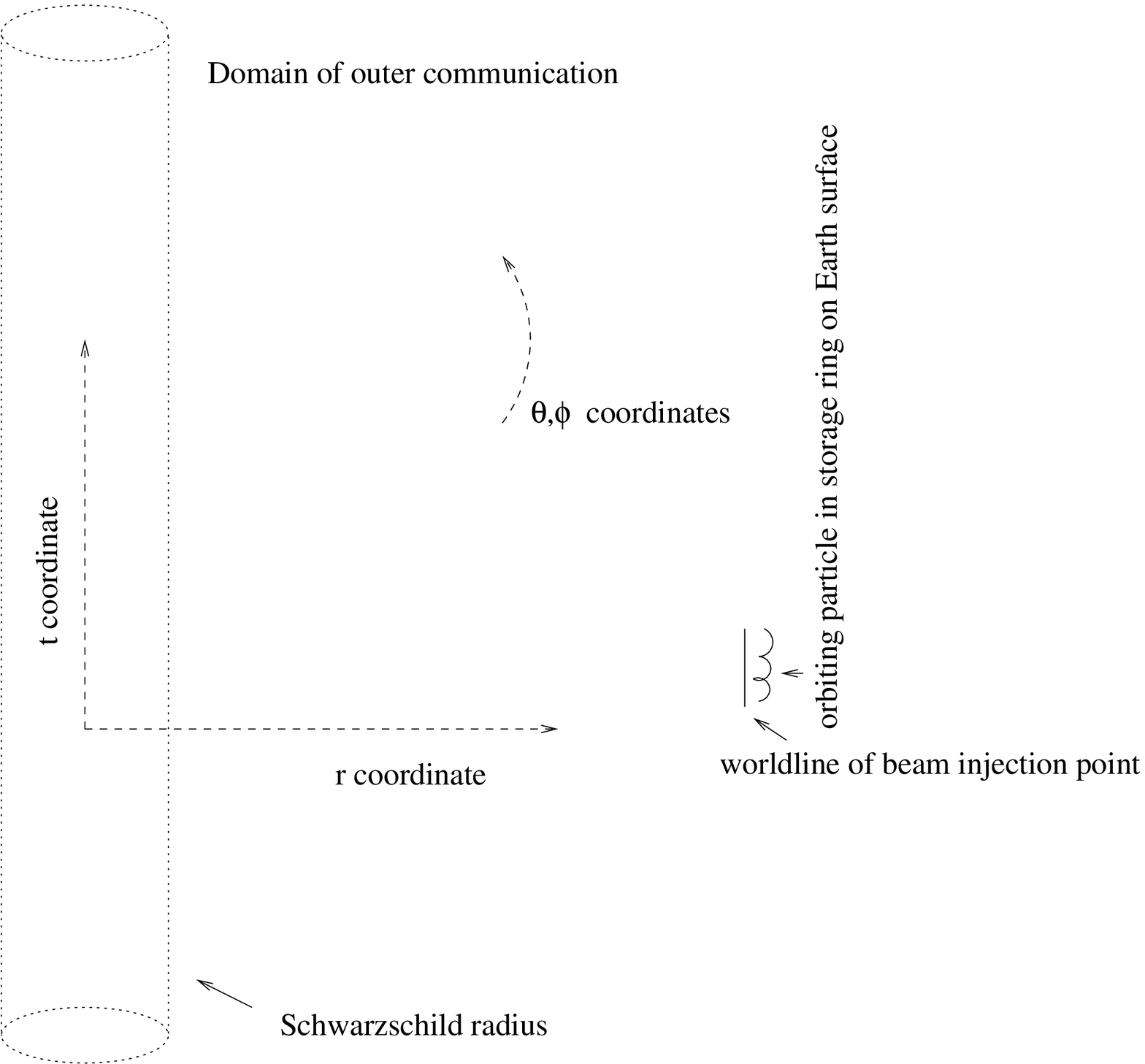}\hspace*{1cm}\includegraphics[width=8cm]{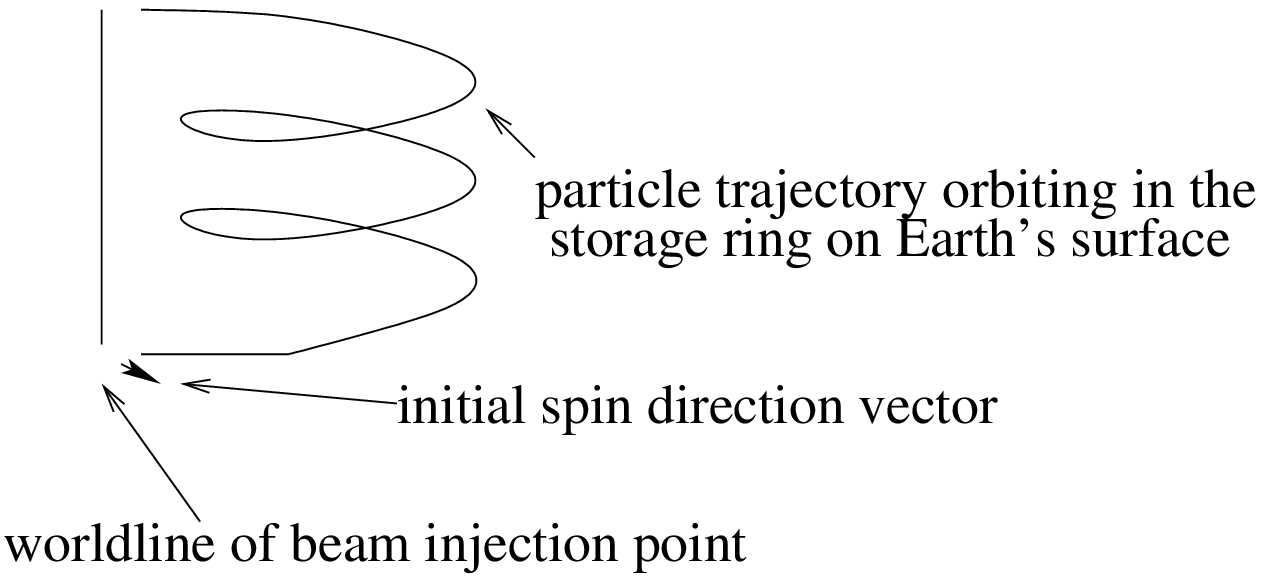}
\end{center}
\caption{The outline of the kinematic setting of the experiment. Top panel: the particle storage ring is sitting 
on the Earth's surface. The radius of the Earth is denoted by $R$, the storage ring radius by 
$L$, and we set the North pole of our spherical coordinates by convention 
to the center of the storage ring. Bottom left panel: illustration of 
the laboratory setting in Earth's Schwarzschild spacetime. Throughout the 
paper the Schwarzschild coordinates $t,r,\vartheta,\varphi$ are used. 
Bottom right panel: a zoom of the particle's orbiting trajectory.
The initial spin direction vector is Fermi-Walker transported along the 
worldline of the orbiting particle in Schwarzschild spacetime. The worldline of the 
beam injection point, i.e., the laboratory worldline is also shown, 
along which the initial spin vector can also be Fermi-Walker transported. }
\label{figSch}
\end{figure}

The kinematic setting of the experiment is outlined in Fig.~\ref{figSch}. The 
gravitational field of the Earth is modelled by a Schwarzschild metric 
with $r_{S}$ being the corresponding Schwarzschild radius, $r_{S}=\frac{2MG}{c^{2}}$. 
The non-sphericity of the Earth as well as its rotation is neglected. 
We use the standard Schwarzschild coordinates $t,r,\vartheta,\varphi$, and thus the 
components of the Schwarzschild metric read as:
\begin{eqnarray}
 g_{\mathsf{a}\mathsf{b}}(t,r,\vartheta,\varphi) & = & \left(\begin{array}{cccc} 1-\frac{r_{S}}{r} & 0 & 0 & 0 \cr 0 & -\frac{1}{1-\frac{r_{S}}{r}} & 0 & 0 \cr 0 & 0 & -r^{2} & 0 \cr 0 & 0 & 0 & -r^{2}\sin^{2}\vartheta \cr\end{array}\right).
\label{eqSchMetric}
\end{eqnarray}
In such coordinates, the Earth's surface is at an $r=\mathrm{const}$ level-surface, we denote this radius by $R$. 
By convention, the North pole of the spherical coordinates is adjusted 
such that it corresponds to the central axis of the storage ring, i.e., this 
axis is at $\vartheta=0$. The entire storage ring is located at an 
$r=R$, $\vartheta=\mathrm{const}$ surface, where the corresponding $\vartheta$ 
coordinate value is denoted by $\Theta$. The radius of the storage ring 
is then $L=R\sin\Theta$. 
Throughout the paper, the 
coordinate indices are denoted by fonts like $\mathsf{a},\mathsf{b},\mathsf{c},\dots$ 
and take their value from the index set $\{0,1,2,3\}$. Occasionally, 
the alternative notation $\{t,r,\vartheta,\varphi\}$ is used as 
equivalent symbols for the indices $\{0,1,2,3\}$. Moreover, we will also use the Penrose abstract indices \cite{wald1984}, 
with index symbol fonts like $a,b,c,\dots$ in order to aid the notation 
of various tensorial trace expressions in a coordinate independent way.

The trajectory of the orbiting particle inside the storage ring on the Earth's 
surface is described by a worldline $t\mapsto\gamma_{\omega}(t)$ with coordinate components
\begin{eqnarray}
\gamma_{\omega}{}^{\mathsf{a}}(t) & = & \left(\begin{array}{c} t \cr R \cr \Theta \cr \omega\sqrt{1-\frac{r_{S}}{R}}\, t \;\mathrm{mod}\; 2\pi \end{array}\right),
\label{eqWorldline}
\end{eqnarray}
where for convenience the worldline is parameterized by the Killing time $t$ 
and not with its proper time.  Here, $\omega$ denotes the circular frequency 
of the orbiting particle trajectory, in terms of the proper time of the laboratory system. 
It is seen that the particles are assumed to be orbiting on a closed 
circular trajectory, i.e., a beam balanced against falling towards the Earth is assumed. 
This is justified by the fact that according to \cite{mane2005}
an electrostatic beam focusing optics is used in the g-2 experimental setup, 
which is resting on the surface of the Earth, together with the storage ring. 
The initial spin direction vector at $t=0$ is a unit pseudolength spacelike vector, orthogonal 
to the curve $t\mapsto\gamma_{\omega}(t)$.
The amount of precession can be quantified via also evolving the initial 
spin direction vector along the worldline of 
the beam injection point of the storage ring (laboratory observer), 
described by the curve $t\mapsto\gamma_{0}(t)$ having coordinate components
\begin{eqnarray}
\gamma_{0}{}^{\mathsf{a}}(t) & = & \left(\begin{array}{c} t \cr R \cr \Theta \cr 0 \end{array}\right).
\label{eqWorldlineRef}
\end{eqnarray}
The worldlines $t\mapsto\gamma_{\omega}(t)$ and $t\mapsto\gamma_{0}(t)$ intersect 
at each full revolution, i.e., at each $t=n\frac{2\pi}{\omega \sqrt{1-\frac{r_{S}}{R}}}$, with $n$ being any non-negative integer. 
In these intersection points the propagated spin direction vectors 
can be eventually compared. 
The unit tangent vector fields, i.e., the four velocity fields of these curves are 
$t\mapsto u_{\omega}{}^{\mathsf{a}}(t):=\frac{1}{\Lambda_{\omega}(t)}\dot{\gamma}_{\omega}{}^{\mathsf{a}}(t)$ and 
$t\mapsto u_{0}{}^{\mathsf{a}}(t):=\frac{1}{\Lambda_{0}(t)}\dot{\gamma}_{0}{}^{\mathsf{a}}(t)$, with 
$\Lambda_{\omega}:=\sqrt{g_{\mathsf{a}\mathsf{b}}\dot{\gamma}_{\omega}{}^{\mathsf{a}}\dot{\gamma}_{\omega}{}^{\mathsf{b}}}$ and 
$\Lambda_{0}:=\sqrt{g_{\mathsf{a}\mathsf{b}}\dot{\gamma}_{0}{}^{\mathsf{a}}\dot{\gamma}_{0}{}^{\mathsf{b}}}$, respectively.

In order to evaluate the spin direction vector along any point of the worldline 
$t\mapsto\gamma_{\omega}(t)$ or $t\mapsto\gamma_{0}(t)$, it needs to be transported along 
the pertinent trajectories. This is described by the \emph{Fermi-Walker transport},
i.e., by the \emph{relativistic gyroscopic transport} \cite{hawking1973}. 
Let $u^{d}$ be a future directed unit timelike vector field, then the \emph{Fermi-Walker derivative} of a 
vector field $w^{b}$ along $u^{d}$ is defined as:
\begin{eqnarray}
 D^{F}_{u}w^{b} & := & u^{d}\nabla_{d}w^{b} \,+\, g_{ac}w^{a}u^{b}u^{d}\nabla_{d}u^{c} \,-\, g_{ac}w^{a}u^{c}u^{d}\nabla_{d}u^{b},
\label{eqFWD}
\end{eqnarray}
where $\nabla_{d}$ denotes the Levi-Civita covariant derivation associated to the metric 
$g_{ab}$. The Fermi-Walker derivative is distinguished by the fact that 
$D^{F}_{u}u^{b}=0$ holds, as well 
as the property that for any two vector field $w^{b}$ and $v^{b}$ satisfying 
$D^{F}_{u}w^{b}=0$ and $D^{F}_{u}v^{b}=0$, the identity $u^{a}\nabla_{a}\left(g_{bc}w^{b}v^{c}\right)=0$ holds. 
In particular, whenever one has $D^{F}_{u}w^{b}=0$, then also $u^{a}\nabla_{a}\left(g_{bc}u^{b}w^{c}\right)=0$ 
and $u^{a}\nabla_{a}\left(g_{bc}w^{b}w^{c}\right)=0$ hold. 
A vector field $w^{b}$ is said to be Fermi-Walker transported along the integral 
curves of a future directed timelike unit vector field $u^{d}$, whenever the equation 
\begin{eqnarray}
 D^{F}_{u}w^{b} & = & 0
\end{eqnarray}
is satisfied, which is just the relativistic gyroscope equation \cite{grav, everitt2011}. The rationale 
behind considering the Fermi-Walker transport as a relativistic model 
of the gyroscope evolution is that for the transport of a vector field $w^{b}$ 
along a timelike curve with future directed unit tangent vector field 
$u^{d}$, the initial constraints
\begin{eqnarray}
 g_{ab}u^{a}u^{b} & = & 1,\cr
 g_{ab}w^{a}w^{b} & = & -1,\cr
 g_{ab}u^{a}w^{b} & = & 0
\end{eqnarray}
are conserved during evolution, and no artificial vorticity is added. Note, that physically the spin vector has constant 
pseudolength and is always perpendicular to the worldline of the particle, 
and this constraint needs to be preserved throughout the evolution. Also note, 
 that intuitively the Fermi-Walker transport can be regarded as the parallel 
transport of a rigid orthonormal frame along a unit timelike vector field, the 
timelike element of the frame coinciding to that of the transporting vector field.

Whenever an electromagnetic field $F_{ab}$ is also present, the charged 
particles with spin are governed by the equations of motion
\begin{eqnarray}
 u^{a}\nabla_{a}u^{b} & = & -\frac{q}{m}\,g^{bc}\,F_{cd}\,u^{d}, \cr
 D^{F}_{u}w^{b} & = & -\frac{\mu}{s}\,\left(g^{bc}\,F_{cd} - u^{b}\,u^{c}\,F_{cd} - g^{bc}\,F_{ce}\,u^{e}\,u^{f}\,g_{fd} \right)\,w^{d},
\label{eqNewtonBMT}
\end{eqnarray}
where the first equation is the relativistic Newton equation with the electromagnetic 
force, and the second equation is the 
\emph{Bargmann-Michel-Telegdi (BMT) equation} \cite{conte1996, jackson1999}. Here, $m$ denotes the particle 
mass, $q$ denotes the particle charge, $\mu$ denotes the magnetic moment 
of the particle, and $s$ denotes the spin magnitude 
($s=\frac{1}{2},1,\frac{3}{2},2,\dots$ for particles), 
while $u^{a}$ is the four velocity of the particle and 
$w^{b}$ is the spin direction vector of the particle. 

Let us note that for charged spinning objects with non-trivial internal 
structure  the BMT equation cannot directly be applied. Instead it can be 
regarded as a special limiting case of the so-called electromagnetically 
extended Mathisson-Papapetrou-Dixon (MPD) equations 
\cite{mathisson37, papapetrou51, dixon64, dixon65, deriglazov2017}, 
which describe in the pole-dipole approximation the motion of a charged 
spinning body on a curved spacetime in the presence of electromagnetic field. 
When the electromagnetic dipole moment tensor is taken to be proportional to 
the spin tensor, and the curvature effects and the second order spin effects 
are all neglected, and after introducing a spin supplementary 
condition\footnote{It is necessary to introduce spin 
supplementary conditions (SSCs) as the MPD equations are not closed. One 
can use, e.g., the Mathisson-Pirani~\cite{mathisson37, pirani56} or the 
Tulczyjew-Dixon \cite{dixon64, tulczyjew59} SSCs.}, the BMT equation is 
obtained. 
In the present context the treated objects are charged particles with spin 
without relevant internal structure, and therefore the BMT equation can be 
safely considered to be enough to describe the spin propagation, and is 
indeed used for the engineering design of accelerator facilities for 
spin-polarized particles.

In Sections~\ref{secFWtransport}, \ref{secRelFWtransport}, \ref{secEvaluationT} 
merely the Fermi-Walker transport $D^{F}_{u}w^{b}=0$, i.e., 
the gyroscopic kinematics of the spin direction vector along the worldlines 
Eq.(\ref{eqWorldline}) and Eq.(\ref{eqWorldlineRef}) will be studied in 
order to extract the Thomas precession over a Schwarzschild background. 
These results apply to any forced circular motion over a Schwarzschild background. 
Following that, in Sections~\ref{secEldin}, \ref{secAbsLarmor}, \ref{secRelLarmor}, \ref{secEvaluationLarmor} 
the GR modifications to the electromagnetic (Larmor) precession is quantified, 
which contributes in addition when the forced circular orbit is achieved via 
an electromagnetic field acting on a charged particle.

\section{The absolute Fermi-Walker transport of four vectors}
\label{secFWtransport}

The Fermi-Walker transport differential equation $D^{F}_{\frac{1}{\Lambda}\dot{\gamma}}w=0$ 
of a vector field $w$ along a curve 
$\lambda\mapsto\gamma(\lambda)$ reads in components as
\begin{eqnarray}
 D^{F}_{\frac{1}{\Lambda}\dot{\gamma}}w^{\mathsf{b}}(\gamma(\lambda))  =  \cr
  \frac{1}{\Lambda(\lambda)}\frac{\mathrm{d}}{\mathrm{d}\lambda} w^{\mathsf{b}}(\gamma(\lambda)) 
\;+\; \frac{1}{\Lambda(\lambda)}\dot{\gamma}^{\mathsf{d}}(\lambda)\,\Gamma_{\mathsf{d}\mathsf{c}}^{\mathsf{b}}(\gamma(\lambda))\,w^{\mathsf{c}}(\gamma(\lambda)) \cr
  \;+\; \frac{1}{\Lambda(\lambda)}g_{\mathsf{a}\mathsf{c}}(\gamma(\lambda))w^{\mathsf{a}}(\gamma(\lambda))\frac{1}{\Lambda^{2}(\lambda)}\dot{\gamma}^{\mathsf{b}}(\lambda)\frac{\mathrm{d}}{\mathrm{d}\lambda}\dot{\gamma}^{\mathsf{c}}(\lambda) \cr
  \;+\; \frac{1}{\Lambda(\lambda)}g_{\mathsf{a}\mathsf{c}}(\gamma(\lambda))w^{\mathsf{a}}(\gamma(\lambda))\frac{1}{\Lambda^{2}(\lambda)}\dot{\gamma}^{\mathsf{b}}(\lambda)\dot{\gamma}^{\mathsf{d}}(\lambda)\Gamma_{\mathsf{d}\mathsf{e}}^{\mathsf{c}}(\gamma(\lambda))\dot{\gamma}^{\mathsf{e}}(\lambda) \cr
  \;-\; \frac{1}{\Lambda(\lambda)}g_{\mathsf{a}\mathsf{c}}(\gamma(\lambda))w^{\mathsf{a}}(\gamma(\lambda))\frac{1}{\Lambda^{2}(\lambda)}\dot{\gamma}^{\mathsf{c}}(\lambda)\frac{\mathrm{d}}{\mathrm{d}\lambda}\dot{\gamma}^{\mathsf{b}}(\lambda) \cr
  \;-\; \frac{1}{\Lambda(\lambda)}g_{\mathsf{a}\mathsf{c}}(\gamma(\lambda))w^{\mathsf{a}}(\gamma(\lambda))\frac{1}{\Lambda^{2}(\lambda)}\dot{\gamma}^{\mathsf{c}}(\lambda)\dot{\gamma}^{\mathsf{d}}(\lambda)\Gamma_{\mathsf{d}\mathsf{e}}^{\mathsf{b}}(\gamma(\lambda))\dot{\gamma}^{\mathsf{e}}(\lambda) \cr
 =  0 \qquad(\lambda\in\R),
\label{eqFWInCoords}
\end{eqnarray}
where $\Gamma_{\mathsf{b}\mathsf{c}}^{\mathsf{a}}$ denotes the Christoffel symbols 
in the used coordinates, and
\begin{eqnarray}
 \lambda\mapsto \Lambda(\lambda) & := & \sqrt{g_{\mathsf{a}\mathsf{b}}(\gamma(\lambda))\dot{\gamma}^{\mathsf{a}}(\lambda)\dot{\gamma}^{\mathsf{b}}(\lambda)}
\end{eqnarray}
is the pseudolength function of the tangent vector field $\lambda\mapsto\dot{\gamma}^{\mathsf{a}}(\lambda)$, 
and $\dot{()}$ denotes derivative against the curve parameter $\lambda$. 
In our calculations, for convenience reasons, we use the Killing time $t$ as 
the parameter of the worldline curves.

In order to calculate Fermi-Walker transported vector fields 
$p\mapsto w_{\omega}{}^{\mathsf{a}}(p)$ and $p\mapsto w_{0}{}^{\mathsf{a}}(p)$ 
along the curves $t\mapsto\gamma_{\omega}(t)$ and $t\mapsto\gamma_{0}(t)$, we introduce the vector valued functions 
$\tilde{w}_{\omega}{}^{\mathsf{a}}(t):=\left(w_{\omega}{}^{\mathsf{a}}\circ \gamma_{\omega}\right)(t)$ and 
$\tilde{w}_{0}{}^{\mathsf{a}}(t):=\left(w_{0}{}^{\mathsf{a}}\circ \gamma_{0}\right)(t)$. 
One should note that the coordinate components of the tangent vectors
\begin{eqnarray}
\dot{\gamma}_{\omega}{}^{\mathsf{a}}(t)  =  \left(\begin{array}{c} 1 \cr 0 \cr 0 \cr \omega\sqrt{1-\frac{r_{S}}{R}} \end{array}\right), \; \; \;
\dot{\gamma}_{0}{}^{\mathsf{a}}(t)  =  \left(\begin{array}{c} 1 \cr 0 \cr 0 \cr 0 \end{array}\right)
\label{eqDotGamma}
\end{eqnarray}
of the curves Eq.(\ref{eqWorldline}) and Eq.(\ref{eqWorldlineRef}) do not 
depend on Killing time, i.e., $\frac{\mathrm{d}}{\mathrm{d}t}\dot{\gamma}_{\omega}{}^{\mathsf{a}}(t)=0$ and $\frac{\mathrm{d}}{\mathrm{d}t}\dot{\gamma}_{0}{}^{\mathsf{a}}(t)=0$ hold. 
Using this, our Fermi-Walker transport equations $\Lambda_{\omega}\,D^{F}_{\frac{1}{\Lambda_{\omega}}\dot{\gamma}_{\omega}}w_{\omega}{}^{\mathsf{a}}=0$ and 
$\Lambda_{0}\,D^{F}_{\frac{1}{\Lambda_{0}}\dot{\gamma}_{0}}w_{0}{}^{\mathsf{a}}=0$ simplify as
\begin{eqnarray}
 \frac{\mathrm{d}}{\mathrm{d}t} \tilde{w}_{\omega}{}^{\mathsf{b}}(t) 
\;+\; \dot{\gamma}_{\omega}{}^{\mathsf{d}}(t)\,\Gamma_{\mathsf{d}\mathsf{c}}^{\mathsf{b}}(\gamma_{\omega}(t))\,\tilde{w}_{\omega}{}^{\mathsf{c}}(t) & & \cr
\;+\; g_{\mathsf{a}\mathsf{c}}(\gamma_{\omega}(t))\tilde{w}_{\omega}{}^{\mathsf{a}}(t)\frac{1}{\Lambda^{2}(t)}\dot{\gamma}_{\omega}{}^{\mathsf{b}}(t)\dot{\gamma}_{\omega}{}^{\mathsf{d}}(t)\Gamma_{\mathsf{d}\mathsf{e}}^{\mathsf{c}}(\gamma_{\omega}(t))\dot{\gamma}_{\omega}{}^{\mathsf{e}}(t) & & \cr
\;-\; g_{\mathsf{a}\mathsf{c}}(\gamma_{\omega}(t))\tilde{w}_{\omega}{}^{\mathsf{a}}(t)\frac{1}{\Lambda^{2}(t)}\dot{\gamma}_{\omega}{}^{\mathsf{c}}(t)\dot{\gamma}_{\omega}{}^{\mathsf{d}}(t)\Gamma_{\mathsf{d}\mathsf{e}}^{\mathsf{b}}(\gamma_{\omega}(t))\dot{\gamma}_{\omega}{}^{\mathsf{e}}(t) 
 & = & 0, \cr
 \frac{\mathrm{d}}{\mathrm{d}t} \tilde{w}_{0}{}^{\mathsf{b}}(t) 
\;+\; \dot{\gamma}_{0}{}^{\mathsf{d}}(t)\,\Gamma_{\mathsf{d}\mathsf{c}}^{\mathsf{b}}(\gamma_{0}(t))\,\tilde{w}_{0}{}^{\mathsf{c}}(t) & & \cr
\;+\; g_{\mathsf{a}\mathsf{c}}(\gamma_{0}(t))\tilde{w}_{0}{}^{\mathsf{a}}(t)\frac{1}{\Lambda^{2}(t)}\dot{\gamma}_{0}{}^{\mathsf{b}}(t)\dot{\gamma}_{0}{}^{\mathsf{d}}(t)\Gamma_{\mathsf{d}\mathsf{e}}^{\mathsf{c}}(\gamma_{0}(t))\dot{\gamma}_{0}{}^{\mathsf{e}}(t) & & \cr
\;-\; g_{\mathsf{a}\mathsf{c}}(\gamma_{0}(t))\tilde{w}_{0}{}^{\mathsf{a}}(t)\frac{1}{\Lambda^{2}(t)}\dot{\gamma}_{0}{}^{\mathsf{c}}(t)\dot{\gamma}_{0}{}^{\mathsf{d}}(t)\Gamma_{\mathsf{d}\mathsf{e}}^{\mathsf{b}}(\gamma_{0}(t))\dot{\gamma}_{0}{}^{\mathsf{e}}(t) 
 & = & 0.
\label{eqFWToSolve}
\end{eqnarray}
These linear differential equations need to be solved for the vector valued 
functions $t\mapsto\tilde{w}_{\omega}{}^{\mathsf{a}}(t)$ and 
$t\mapsto\tilde{w}_{0}{}^{\mathsf{a}}(t)$.

In order to solve the transport equations Eq.(\ref{eqFWToSolve}), one 
needs the expressions of the Christoffel symbols over Schwarzschild spacetime 
in our coordinate conventions. The only non-vanishing components at a point 
$t,r,\vartheta,\varphi$ are:
\begin{eqnarray}
 \Gamma_{tt}^{r}(t,r,\vartheta,\varphi) & = & \frac{(r-r_{S})r_{S}}{2r^{3}}, \cr
 \Gamma_{tr}^{t}(t,r,\vartheta,\varphi) & = & \frac{r_{S}}{2r(r-r_{S})}, \cr
 \Gamma_{rr}^{r}(t,r,\vartheta,\varphi) & = & -\frac{r_{S}}{2r(r-r_{S})}, \cr
 \Gamma_{r\vartheta}^{\vartheta}(t,r,\vartheta,\varphi) & = & \frac{1}{r}, \cr
 \Gamma_{r\varphi}^{\varphi}(t,r,\vartheta,\varphi) & = & \frac{1}{r}, \cr
 \Gamma_{\vartheta\vartheta}^{r}(t,r,\vartheta,\varphi) & = & -(r-r_{S}), \cr
 \Gamma_{\vartheta\varphi}^{\varphi}(t,r,\vartheta,\varphi) & = & \frac{\cos\vartheta}{\sin\vartheta}, \cr
 \Gamma_{\varphi\varphi}^{r}(t,r,\vartheta,\varphi) & = & -(r-r_{S})\sin^{2}\vartheta, \cr
 \Gamma_{\varphi\varphi}^{\vartheta}(t,r,\vartheta,\varphi) & = & -\sin\vartheta \cos\vartheta,
\label{eqChr}
\end{eqnarray}
where the index symmetry property $\Gamma_{\mathsf{b}\mathsf{c}}^{\mathsf{a}}=\Gamma_{\mathsf{c}\mathsf{b}}^{\mathsf{a}}$ 
also needs to be taken into account. Observe, that due to the time translational 
and spherical symmetry of the Schwarzschild spacetime, the Christoffel 
symbols $\Gamma_{\mathsf{b}\mathsf{c}}^{\mathsf{a}}$ in our adapted coordinates only have $\vartheta$ 
dependence on $r=\mathrm{const}$ surfaces, i.e., also on the $r=R$ surface of 
the Earth. Since the curves $t\mapsto\gamma_{\omega}(t)$ and $t\mapsto\gamma_{0}(t)$ evolve on the Earth's surface, i.e., on the 
$r=R$ surface, the Christoffel symbol coefficients in Eq.(\ref{eqFWToSolve}) 
can merely have $\vartheta$ dependence along these curves. But since the pertinent 
curves are also $\vartheta=\mathrm{const}$, or more precisely $\vartheta=\Theta$ 
curves, the Christoffel symbol coefficients in Eq.(\ref{eqFWToSolve}) are 
completely constant along these. 
Similarly, the metric tensor components $g_{\mathsf{a}\mathsf{b}}$ are also constants 
along these world lines. 
Moreover, also the vector valued functions $t\mapsto\dot{\gamma}_{\omega}{}^{\mathsf{a}}(t)$ 
and $t\mapsto\dot{\gamma}_{0}{}^{\mathsf{a}}(t)$ are constant. 
All these imply that the homogeneous linear differential equations 
Eq.(\ref{eqFWToSolve}) have constant coefficients, and therefore they can 
be eventually solved relatively easily, by a matrix exponentiation.

In the following, we denote by the symbol 
$\Gamma_{\mathsf{b}\mathsf{c}}^{\mathsf{a}}$ the particular constant value 
of the Schwarzschild Christoffel symbols along the curves $t\mapsto\gamma_{\omega}(t)$ 
or $t\mapsto\gamma_{0}(t)$, in our coordinates. 
Similarly, $g_{\mathsf{a}\mathsf{b}}$ will denote the particular constant 
value of the metric tensor components along these world lines. 
These are obtained by simply substituting the values $r=R$, $\vartheta=\Theta$ 
and any value of $\varphi$ and $t$ into Eq.(\ref{eqChr}) and Eq.(\ref{eqSchMetric}). 
Similarly, the symbol $\dot{\gamma}_{\omega}{}^{\mathsf{a}}$ and $\dot{\gamma}_{0}{}^{\mathsf{a}}$ will denote the 
constant value of the constant vector valued functions Eq.(\ref{eqDotGamma}). 
Also, their pseudolengths are constant, $\Lambda_{\omega}=\sqrt{(1-\frac{r_{S}}{R})(1-\omega^{2}R^{2}\sin^{2}\Theta)}$ and $\Lambda_{0}=\sqrt{1-\frac{r_{S}}{R}}$. 
With these notations, we are left with  homogeneous linear differential equations 
with constant coefficients:
\begin{eqnarray}
 \frac{\mathrm{d}}{\mathrm{d}t} \tilde{w}_{\omega}{}^{\mathsf{b}}(t) 
\;+\; \dot{\gamma}_{\omega}{}^{\mathsf{d}}\,\Gamma_{\mathsf{d}\mathsf{c}}^{\mathsf{b}}\,\tilde{w}_{\omega}{}^{\mathsf{c}}(t) & & \cr
\;+\; g_{\mathsf{c}\mathsf{a}}\tilde{w}_{\omega}{}^{\mathsf{c}}(t)\frac{1}{\Lambda_{\omega}{}^{2}}\dot{\gamma}_{\omega}{}^{\mathsf{b}}\dot{\gamma}_{\omega}{}^{\mathsf{d}}\Gamma_{\mathsf{d}\mathsf{e}}^{\mathsf{a}}\dot{\gamma}_{\omega}{}^{\mathsf{e}} 
\;-\; g_{\mathsf{c}\mathsf{a}}\tilde{w}_{\omega}{}^{\mathsf{c}}(t)\frac{1}{\Lambda_{\omega}{}^{2}}\dot{\gamma}_{\omega}{}^{\mathsf{a}}\dot{\gamma}_{\omega}{}^{\mathsf{d}}\Gamma_{\mathsf{d}\mathsf{e}}^{\mathsf{b}}\dot{\gamma}_{\omega}{}^{\mathsf{e}} 
 & = & 0, \cr
 \frac{\mathrm{d}}{\mathrm{d}t} \tilde{w}_{0}{}^{\mathsf{b}}(t) 
\;+\; \dot{\gamma}_{0}{}^{\mathsf{d}}\,\Gamma_{\mathsf{d}\mathsf{c}}^{\mathsf{b}}\,\tilde{w}_{0}{}^{\mathsf{c}}(t) & & \cr
\;+\; g_{\mathsf{c}\mathsf{a}}\tilde{w}_{0}{}^{\mathsf{c}}(t)\frac{1}{\Lambda_{0}{}^{2}}\dot{\gamma}_{0}{}^{\mathsf{b}}\dot{\gamma}_{0}{}^{\mathsf{d}}\Gamma_{\mathsf{d}\mathsf{e}}^{\mathsf{a}}\dot{\gamma}_{0}{}^{\mathsf{e}} 
\;-\; g_{\mathsf{c}\mathsf{a}}\tilde{w}_{0}{}^{\mathsf{c}}(t)\frac{1}{\Lambda_{0}{}^{2}}\dot{\gamma}_{0}{}^{\mathsf{a}}\dot{\gamma}_{0}{}^{\mathsf{d}}\Gamma_{\mathsf{d}\mathsf{e}}^{\mathsf{b}}\dot{\gamma}_{0}{}^{\mathsf{e}} 
 & = & 0. \cr
 & & 
\end{eqnarray}

Direct evaluation shows that 
$\dot{\gamma}_{0}{}^{\mathsf{d}}\,\Gamma_{\mathsf{d}\mathsf{c}}^{\mathsf{b}} 
+ g_{\mathsf{c}\mathsf{a}}\frac{1}{\Lambda_{0}{}^{2}}\dot{\gamma}_{0}{}^{\mathsf{b}}\dot{\gamma}_{0}{}^{\mathsf{d}}\Gamma_{\mathsf{d}\mathsf{e}}^{\mathsf{a}}\dot{\gamma}_{0}{}^{\mathsf{e}} 
- g_{\mathsf{c}\mathsf{a}}\frac{1}{\Lambda_{0}{}^{2}}\dot{\gamma}_{0}{}^{\mathsf{a}}\dot{\gamma}_{0}{}^{\mathsf{d}}\Gamma_{\mathsf{d}\mathsf{e}}^{\mathsf{b}}\dot{\gamma}_{0}{}^{\mathsf{e}} =0 $
holds, and thus any Fermi-Walker transported vector field $t\mapsto \tilde{v}_{0}{}^{\mathsf{b}}(t)$ along 
the curve $t\mapsto\gamma_{0}(t)$ satisfies $\frac{\mathrm{d}}{\mathrm{d}t}\tilde{v}_{0}{}^{\mathsf{b}}(t)=0$ 
for all $t\in\R$. It also means that the Fermi-Walker derivative along 
$\frac{1}{\Lambda_{0}}\dot{\gamma}_{0}$ is proportional to the Lie derivative 
against the Killing time translation vector field $\partial_{t}$. 
Taking this into account, our pair of differential equations simplify as
\begin{eqnarray}
 \frac{\mathrm{d}}{\mathrm{d}t} \tilde{w}_{\omega}{}^{\mathsf{b}}(t) = \F_{\omega}{}^{\mathsf{b}}{}_{\mathsf{c}}\,\tilde{w}_{\omega}{}^{\mathsf{c}}(t), \cr
 \frac{\mathrm{d}}{\mathrm{d}t} \tilde{w}_{0}{}^{\mathsf{b}}(t) = 0,
\end{eqnarray}
with 
\begin{eqnarray}
 \F_{\omega}{}^{\mathsf{b}}{}_{\mathsf{c}} := 
 -\dot{\gamma}_{\omega}{}^{\mathsf{d}}\,\Gamma_{\mathsf{d}\mathsf{c}}^{\mathsf{b}}
 - g_{\mathsf{c}\mathsf{a}}\frac{1}{\Lambda_{\omega}{}^{2}}\dot{\gamma}_{\omega}{}^{\mathsf{b}}\dot{\gamma}_{\omega}{}^{\mathsf{d}}\Gamma_{\mathsf{d}\mathsf{e}}^{\mathsf{a}}\dot{\gamma}_{\omega}{}^{\mathsf{e}} 
 + g_{\mathsf{c}\mathsf{a}}\frac{1}{\Lambda_{\omega}{}^{2}}\dot{\gamma}_{\omega}{}^{\mathsf{a}}\dot{\gamma}_{\omega}{}^{\mathsf{d}}\Gamma_{\mathsf{d}\mathsf{e}}^{\mathsf{b}}\dot{\gamma}_{\omega}{}^{\mathsf{e}}
\label{eqFDef}
\end{eqnarray}
being the \emph{Fermi-Walker transport tensor}. The index pulled up version 
$\F_{\omega}{}^{\mathsf{b}}{}_{\mathsf{c}}g^{\mathsf{c}\mathsf{d}}$ of the 
Fermi-Walker transport tensor can be shown to be antisymmetric by direct substitution. 
Therefore, it describes a Lorentz transformation generator. Moreover, 
$\F_{\omega}{}^{\mathsf{b}}{}_{\mathsf{c}}u_{\omega}{}^{\mathsf{c}}=0$ holds by construction. Therefore, 
the Fermi-Walker transport tensor $\F_{\omega}{}^{\mathsf{b}}{}_{\mathsf{c}}$ 
describes a pure rotation in the space of $u_{\omega}{}^{\mathsf{a}}$-orthogonal vectors, 
called to be the \emph{Thomas rotation}, and describes an absolute, i.e., 
observer independent rotation effect of the spin direction four vector.
The concrete formula for the Fermi-Walker transport tensor is

{
\hspace*{-2.9cm}\begin{minipage}{\textwidth}
\begin{eqnarray}
 \F_{\omega}{}^{\mathsf{b}}{}_{\mathsf{c}} = \frac{\omega \sqrt{1-\frac{r_{S}}{R}} }{1-\omega^{2}L^{2}} \cr
\quad\left(\begin{array}{cccc}
0 & -\omega L\frac{L}{R}\frac{\left(1-\frac{3}{2}\frac{r_{S}}{R}\right)}{\left(1-\frac{r_{S}}{R}\right)^{\frac{3}{2}}} & -\omega L R \frac{\sqrt{1-\left(\frac{L}{R}\right)^{2}}}{\left(1-\frac{r_{S}}{R}\right)^{\frac{1}{2}}} & 0 \cr
-\omega L \frac{L}{R}(1-\frac{3}{2}\frac{r_{S}}{R})\left(1-\frac{r_{S}}{R}\right)^{\frac{1}{2}} & 0 & 0 & L \frac{L}{R}\left(1-\frac{3}{2}\frac{r_{S}}{R}\right) \cr
 & & & \cr
-\omega L \frac{1}{R}\sqrt{1-\left(\frac{L}{R}\right)^{2}}\left(1-\frac{r_{S}}{R}\right)^{\frac{1}{2}} & 0 & 0 & \frac{L}{R}\sqrt{1-\left(\frac{L}{R}\right)^{2}} \cr
0 & -\frac{1}{R}\frac{\left(1-\frac{3}{2}\frac{r_{S}}{R}\right)}{\left(1-\frac{r_{S}}{R}\right)} & -\frac{R}{L}\sqrt{1-\left(\frac{L}{R}\right)^{2}} & 0 \cr
\end{array}\right)
\cr\cr\cr
\label{eqFComps}
\end{eqnarray}
\end{minipage}
}

\noindent
in our coordinate conventions.

\section{The relative Fermi-Walker transport as seen by the laboratory observer}
\label{secRelFWtransport}

As shown in the previous section, the Fermi-Walker transport of four vectors 
along $t\mapsto\gamma_{\omega}(t)$ is relatively simple notion described by the 
tensor $\F_{\omega}{}^{\mathsf{b}}{}_{\mathsf{c}}$. This needs to be translated 
to the transport of spatial vectors orthogonal to the laboratory observer 
$u_{0}{}^{\mathsf{a}}$, known to be the \emph{Thomas precession}, which is a 
phenomenon also including effects relative to an observer. The procedure for quantifying this effect 
is rather well known already in the special relativistic scenario 
\cite{rindler2006, matolcsi2007}.

\begin{figure}[!h]
\begin{center}
\includegraphics[width=12cm]{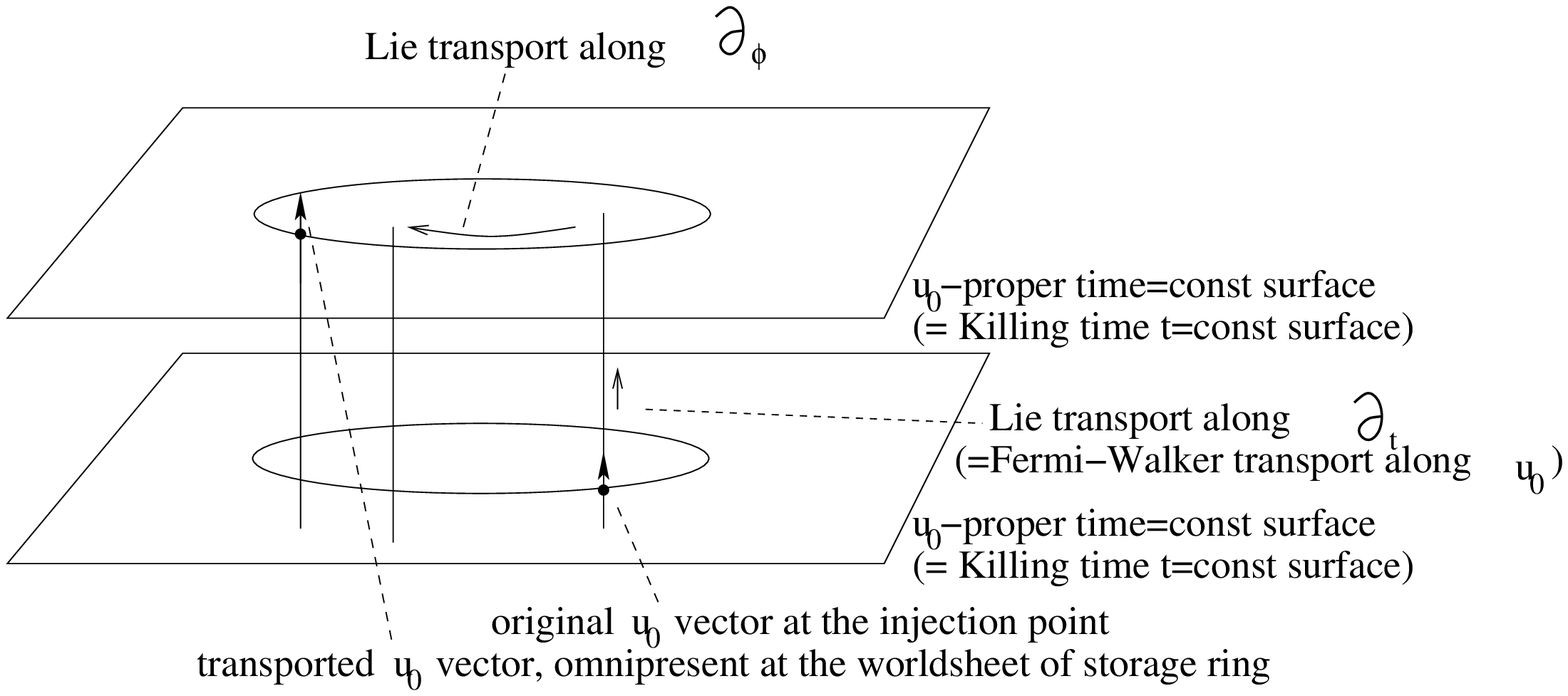}

\vspace*{10mm}
\includegraphics[width=12cm]{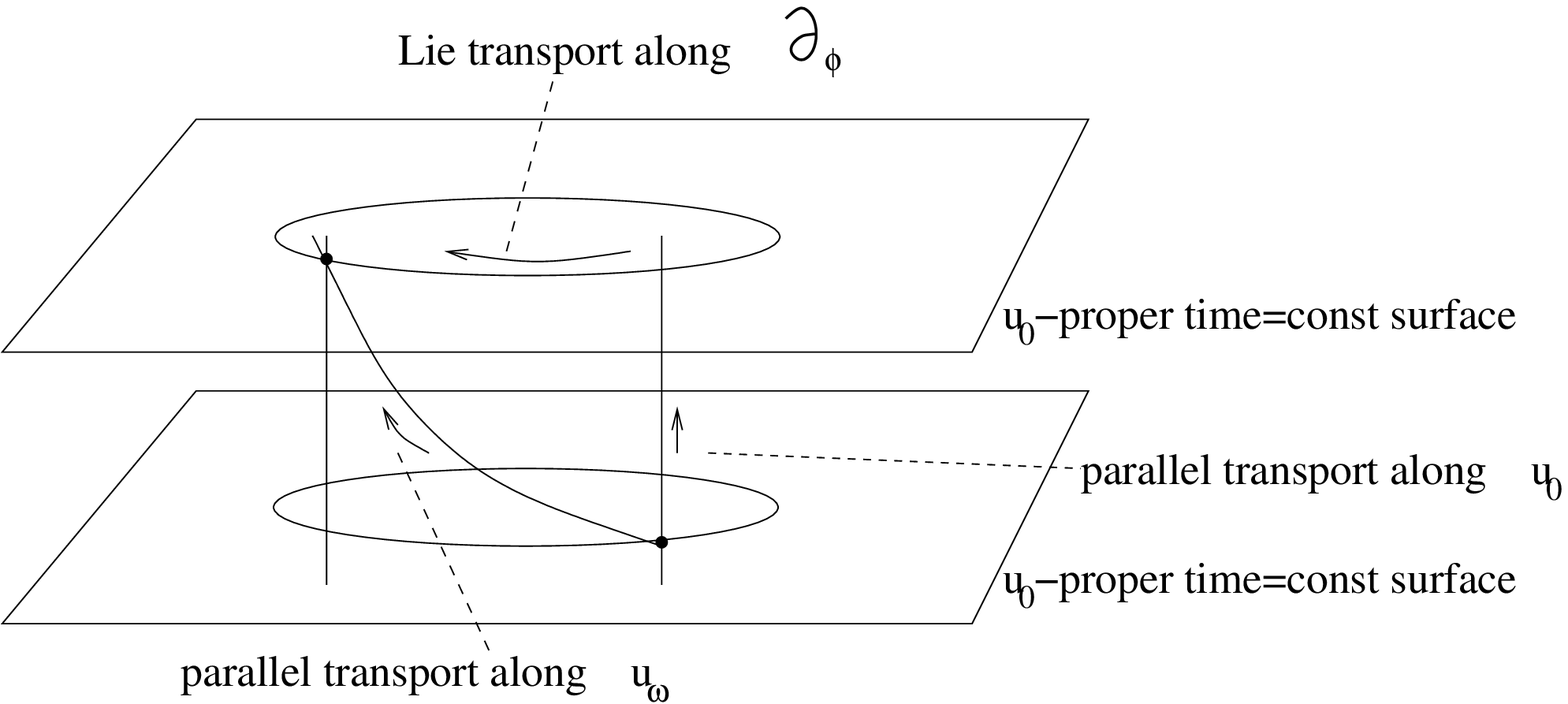}
\end{center}
\caption{Top panel: illustration of how the four velocity vector 
$u_{0}$ of the beam injection point is spread along the worldsheet of the 
storage ring. It is spread via Lie extension, i.e., via requiring $\mathcal{L}_{\partial_{t}}u_{0}=0$ 
and $\mathcal{L}_{\partial_{\varphi}}u_{0}=0$, as described in the text. 
Since it is vorticity-free, by means of Frobenius theorem it can be 
Einstein synchronized by orthogonal surfaces, 
which happen to coincide with Killing time $t=\mathrm{const}$ level surfaces. 
Actually, any vector at an initial point of the pertinent worldsheet can be 
extended to the entire worldsheet via such Lie extension. 
Bottom panel: illustration of how the Einstein synchronized observer $u_{0}$ measures 
evolution of vector fields along the curve $t\mapsto\gamma_{\omega}(t)$ in terms of observer time. 
The evolution of the vector field 
in terms of parallel transport of along $u_{\omega}$ is compared to the 
evolution of the Lie extended initial vector in terms of parallel transport 
along $u_{0}$ and subsequent Lie transport along $\partial_{\varphi}$ in 
the matching comparison spacetime point.}
\label{figTimeDerivative}
\end{figure}

Recall that the worldline of the beam injection point in the laboratory is 
the curve $t\mapsto\gamma_{0}(t)$ with a four velocity vector $u_{0}$, described by Eq.(\ref{eqWorldlineRef}). Let 
us consider such a curve in each point of the storage ring. In other words: 
take the initial $u_{0}$ vector, and extend it via requiring 
$\mathcal{L}_{\partial_{t}}u_{0}=0$ to all $t$, defining the four velocity field of the curve 
$t\mapsto\gamma_{0}(t)$. It will obey the Fermi-Walker transport equation 
$D^{F}_{u_{0}}u_{0}=\frac{1}{\Lambda_{0}}\mathcal{L}_{\partial_{t}}u_{0}=0$ 
along itself. Then, extend it via the Lie transport 
$\mathcal{L}_{\partial_{\varphi}}$ to any point of the storage ring world sheet. 
This $u_{0}$ vector field will have a family of integral curves 
\begin{eqnarray}
 t\mapsto\gamma_{0,\phi}(t):=\left(\begin{array}{c}t \cr R \cr \Theta \cr \phi \cr \end{array}\right)
\end{eqnarray}
indexed by $\phi\in[0,2\pi[$. These will be the worldlines 
of the laboratory observer. Similarly, to Eq.(\ref{eqWorldlineRef}), these will have the 
tangent vector field 
\begin{eqnarray}
 \dot{\gamma}_{0}{}^{\mathsf{a}}(t,R,\Theta,\varphi):=\left(\begin{array}{c}1 \cr 0 \cr 0 \cr 0 \cr \end{array}\right),
\end{eqnarray}
and will have corresponding unit tangent vector field, i.e., four velocity 
$u_{0}{}^{\mathsf{a}}(t,R,\Theta,\varphi):=\frac{1}{\Lambda_{0}(t,R,\Theta,\varphi)}\dot{\gamma}_{0}{}^{\mathsf{a}}(t,R,\Theta,\varphi)$ 
with $\Lambda_{0}:=\sqrt{g_{\mathsf{a}\mathsf{b}}\dot{\gamma}_{0}{}^{\mathsf{a}}\dot{\gamma}_{0}{}^{\mathsf{b}}}=\sqrt{1-\frac{r_{S}}{R}}$. 
By this construction, the observer vector field 
$u_{0}$ is present at each point of the storage ring world sheet as shown 
in the top panel of Fig.~\ref{figTimeDerivative}, with the property 
$\mathcal{L}_{\partial_{t}}u_{0}=0$, $\mathcal{L}_{\partial_{\varphi}}u_{0}=0$. 
Actually, any vector $v$ at the initial spacetime point can be spread as a reference vector to 
any point of the storage ring worldsheet, using this ``Lie extension''
$\mathcal{L}_{\partial_{t}}v=0$, $\mathcal{L}_{\partial_{\varphi}}v=0$. 
Since this spread vector field $u_{0}$ is vorticity-free, by means of 
Frobenius theorem it can be Einstein 
synchronized with orthogonal surfaces. These happen to coincide with the 
Killing time $t=\mathrm{const}$ surfaces. 
The Einstein synchronized observer $u_{0}$ observes $u_{0}$-time evolution of 
vector fields along the curve $t\mapsto\gamma_{\omega}(t)$ via first 
spreading the initial vector using the above Lie extension as a reference, 
and then comparing the parallel transport evolution of the vector field along 
$u_{\omega}$ to the evolution of the Lie extended spread reference 
vector field along the $u_{0}$ parallel transport and subsequent 
$\partial_{\varphi}$ Lie transport, in order to match the comparison spacetime point. 
This is illustrated in the bottom panel of Fig.~\ref{figTimeDerivative}. 
As a consequence, the covariant $u_{0}$-time derivative of vector fields $v^{a}$ along 
$t\mapsto\gamma_{\omega}(t)$ formally can be written as 
$\left(v^{a}\right)':=\frac{\Lambda_{\omega}}{\Lambda_{0}}u_{\omega}{}^{d}\nabla_{d}v^{a}-u_{0}{}^{d}\nabla_{d}\check{v}^{a}-\omega\mathcal{L}_{\partial_{\varphi}}\check{v}^{a}$, 
where $\check{v}^{a}$ denotes the Lie extended vector field of the vector 
$v^{a}$ at the given point of the curve, in order to make sense of the formula. 
In terms of coordinate components, this is described by 
\begin{eqnarray}
 \left(v^{\mathsf{a}}\right)'(\gamma_{\omega}(t)) = \cr
\frac{\Lambda_{\omega}}{\Lambda_{0}}\left(\frac{1}{\Lambda_{\omega}}\frac{\mathrm{d}}{\mathrm{d}t}v^{\mathsf{a}}(\gamma_{\omega}(t))+\frac{1}{\Lambda_{\omega}}\dot{\gamma}_{\omega}{}^{\mathsf{b}}\Gamma_{\mathsf{b}\mathsf{c}}^{\mathsf{a}}v^{\mathsf{c}}(\gamma_{\omega}(t))\right)-\frac{1}{\Lambda_{0}}\dot{\gamma}_{0}{}^{\mathsf{b}}\Gamma_{\mathsf{b}\mathsf{c}}^{\mathsf{a}}v^{\mathsf{c}}(\gamma_{\omega}(t))
\label{eqPrime},
\end{eqnarray}
for a vector field $p\mapsto v^{\mathsf{a}}(p)$ along the curve $t\mapsto\gamma_{\omega}(t)$, 
in our coordinate choice.

It is important to recall that the evolving Fermi-Walker transported spin 
direction vector field $w_{\omega}$ is always orthogonal to $u_{\omega}$. 
Let us denote by $E_{u_{\omega}}$ at a point of $t\mapsto\gamma_{\omega}(t)$ 
the subspace of $u_{\omega}$-orthogonal vectors ($u_{\omega}$-space vectors). 
Also, let $E_{u_{0}}$ denote the orthogonal vectors to $u_{0}$ at a point 
of the laboratory observer world sheet.

Take a solution $w_{\omega}{}^{a}$ of the Fermi-Walker transport equation 
$D^{F}_{u_{\omega}}w_{\omega}{}^{a}=0$, 
where the vector $w_{\omega}{}^{a}$ is initially (and thus also eternally) 
$u_{\omega}$-space vector, i.e., resides in $E_{u_{\omega}}$. The Einstein 
synchronized laboratory observer $u_{0}$, at a corresponding spacetime point, 
observes it via Lorentz boosting it 
back to $E_{u_{0}}$. That shall be denoted by $w_{\omega,u_{0}}$, being an 
$u_{0}$-space vector at the same spacetime point. The Lorentz boost at a spacetime 
point from a future directed unit timelike vector $u_{1}$ to an other one $u_{2}$ is given by the formula:
\begin{eqnarray}
 B_{u_{2},u_{1}}{}^{b}{}_{c} & = & \delta^{b}{}_{c} - \frac{(u_{2}{}^{b}+u_{1}{}^{b})(u_{2}{}^{d}+u_{1}{}^{d})g_{dc}}{1+g_{ef}u_{2}{}^{e}u_{1}{}^{f}} + 2u_{2}{}^{b}g_{cd}u_{1}{}^{d}.
\end{eqnarray}
It is uniquely characterized by the following properties:  it is the $g_{ab}$-isometry 
taking $u_{1}$ to $u_{2}$ (and thus $E_{u_{1}}$ to $E_{u_{2}}$) that acts as the identity on the subspace $E_{u_{1}}\cap E_{u_{2}}$. With this 
notation, one has that the original Fermi-Walker transported four vector field 
is described by 
$w_{\omega}{}^{a}=B_{u_{\omega},u_{0}}{}^{a}{}_{b}w_{\omega,u_{0}}{}^{b}$. 
Since that was required to satisfy the Fermi-Walker transport equation, it 
must satisfy
\begin{eqnarray}
 u_{\omega}{}^{d}\nabla_{d}\,(B_{u_{\omega},u_{0}}{}^{a}{}_{b}w_{\omega,u_{0}}{}^{b}) & = & -u_{\omega}{}^{a}(u_{\omega}{}^{d}\nabla_{d}u_{\omega}{}^{c})g_{ce}(B_{u_{\omega},u_{0}}{}^{e}{}_{b}w_{\omega,u_{0}}{}^{b}) \cr
 & & +u_{\omega}{}^{c}(u_{\omega}{}^{d}\nabla_{d}u_{\omega}{}^{a})g_{ce}(B_{u_{\omega},u_{0}}{}^{e}{}_{b}w_{\omega,u_{0}}{}^{b})
\end{eqnarray}
along the curve $t\mapsto\gamma_{\omega}(t)$. Applying now inverse boost $B_{u_{0},u_{\omega}}$, 
i.e., boost from $u_{\omega}$ to $u_{0}$, and using subsequently the Leibniz 
rule for covariant derivation, one infers that
\begin{eqnarray}
 u_{\omega}{}^{d}\nabla_{d}\,w_{\omega,u_{0}}{}^{f} & = & -u_{\omega}{}^{d}\left(B_{u_{0},u_{\omega}}{}^{f}{}_{a}\nabla_{d}B_{u_{\omega},u_{0}}{}^{a}{}_{b}\right)\,w_{\omega,u_{0}}{}^{b} \cr
 & & -B_{u_{0},u_{\omega}}{}^{f}{}_{a}u_{\omega}{}^{a}(u_{\omega}{}^{d}\nabla_{d}u_{\omega}{}^{c})g_{ce}B_{u_{\omega},u_{0}}{}^{e}{}_{b}\,w_{\omega,u_{0}}{}^{b} \cr
 & & +B_{u_{0},u_{\omega}}{}^{f}{}_{a}u_{\omega}{}^{c}(u_{\omega}{}^{d}\nabla_{d}u_{\omega}{}^{a})g_{ce}B_{u_{\omega},u_{0}}{}^{e}{}_{b}\,w_{\omega,u_{0}}{}^{b}
\end{eqnarray}
must be satisfied. Using this and Eq.(\ref{eqPrime}), the $u_{0}$-time derivative 
of the observed Fermi-Walker transported vector field $w_{\omega,u_{0}}$ can 
be given:
\begin{eqnarray}
 \left(w_{\omega,u_{0}}{}^{\mathsf{f}}\right)' & = & \Phi^{T}_{\omega,u_{0}}{}^{\mathsf{f}}{}_{\mathsf{b}}\,w_{\omega,u_{0}}{}^{\mathsf{b}},
\label{eqThomasMotion}
\end{eqnarray}
with the $u_{0}$-Fermi-Walker transport tensor
\begin{eqnarray}
 \Phi^{T}_{\omega,u_{0}}{}^{\mathsf{f}}{}_{\mathsf{b}} & := & -\frac{1}{\Lambda_{0}}\dot{\gamma}_{0}{}^{\mathsf{d}}\Gamma_{\mathsf{d}\mathsf{b}}^{\mathsf{f}} \cr
 & & -\frac{1}{\Lambda_{0}}\dot{\gamma}_{\omega}{}^{\mathsf{d}}\left(B_{u_{0},u_{\omega}}{}^{\mathsf{f}}{}_{\mathsf{a}}\nabla_{\mathsf{d}}B_{u_{\omega},u_{0}}{}^{\mathsf{a}}{}_{\mathsf{b}}\right) \cr
 & & -\frac{1}{\Lambda_{0}}B_{u_{0},u_{\omega}}{}^{\mathsf{f}}{}_{\mathsf{a}}\dot{\gamma}_{\omega}{}^{\mathsf{a}}(u_{\omega}{}^{\mathsf{d}}\nabla_{\mathsf{d}}u_{\omega}{}^{\mathsf{c}})g_{\mathsf{c}\mathsf{e}}B_{u_{\omega},u_{0}}{}^{\mathsf{e}}{}_{\mathsf{b}} \cr
 & & +\frac{1}{\Lambda_{0}}B_{u_{0},u_{\omega}}{}^{\mathsf{f}}{}_{\mathsf{a}}\dot{\gamma}_{\omega}{}^{\mathsf{c}}(u_{\omega}{}^{\mathsf{d}}\nabla_{\mathsf{d}}u_{\omega}{}^{\mathsf{a}})g_{\mathsf{c}\mathsf{e}}B_{u_{\omega},u_{0}}{}^{\mathsf{e}}{}_{\mathsf{b}}.
\end{eqnarray}
Using now the fact that we took special coordinates such that the coordinate 
components of $u_{\omega}{}^{\mathsf{a}}$, $u_{0}{}^{\mathsf{a}}$ and 
$g_{\mathsf{a}\mathsf{b}}$ are 
constant, we get an explicit form for the coordinate components
\begin{eqnarray}
 \Phi^{T}_{\omega,u_{0}}{}^{\mathsf{f}}{}_{\mathsf{b}} & = & 
 -\frac{1}{\Lambda_{0}}\dot{\gamma}_{0}{}^{\mathsf{d}}\Gamma_{\mathsf{d}\mathsf{b}}^{\mathsf{f}}+\frac{1}{\Lambda_{0}}\dot{\gamma}_{\omega}^{\mathsf{d}}\Gamma_{\mathsf{d}\mathsf{b}}^{\mathsf{f}} + \frac{1}{\Lambda_{0}}B_{u_{0},u_{\omega}}{}^{\mathsf{f}}{}_{\mathsf{a}}\,\F_{\omega}{}^{\mathsf{a}}{}_{\mathsf{e}}\,B_{u_{\omega},u_{0}}{}^{\mathsf{e}}{}_{\mathsf{b}}.
\end{eqnarray}
By direct substitution it is seen that the index pulled 
up version $\Phi^{T}_{\omega,u_{0}}{}^{\mathsf{f}}{}_{\mathsf{b}}g^{\mathsf{b}\mathsf{c}}$ 
is antisymmetric, and therefore corresponds to a Lorentz transformation 
generator. Also, it is seen that 
$\Phi^{T}_{\omega,u_{0}}{}^{\mathsf{f}}{}_{\mathsf{b}}u_{0}{}^{\mathsf{b}}=0$, and 
therefore it is an $u_{0}$-rotation generator, called to be the 
\emph{Thomas precession}, which includes the relative observer effects as well.
The concrete coordinate components of the Thomas precession tensor 
 is
\begin{eqnarray}
\Phi^{T}_{\omega,u_{0}}{}^{\mathsf{a}}{}_{\mathsf{b}} & = & \left(\begin{array}{cccc}
0 & 0 & 0 & 0 \cr
0 & 0 & 0 & \Phi^{T}_{\omega,u_{0}}{}^{r}{}_{\varphi} \cr
0 & 0 & 0 & \Phi^{T}_{\omega,u_{0}}{}^{\vartheta}{}_{\varphi} \cr
0 & \Phi^{T}_{\omega,u_{0}}{}^{\varphi}{}_{r} & \Phi^{T}_{\omega,u_{0}}{}^{\varphi}{}_{\vartheta} & 0 \cr
\end{array}\right), \cr
  \mathrm{with}& & \cr
 \Phi^{T}_{\omega,u_{0}}{}^{\varphi}{}_{r} & = & -\Phi^{T}_{\omega,u_{0}}{}^{r}{}_{\varphi}\,\frac{g^{\varphi\varphi}}{g^{rr}}, \cr
 \Phi^{T}_{\omega,u_{0}}{}^{\varphi}{}_{\vartheta} & = & -\Phi^{T}_{\omega,u_{0}}{}^{\vartheta}{}_{\varphi}\,\frac{g^{\varphi\varphi}}{g^{\vartheta\vartheta}}, \cr
\Phi^{T}_{\omega,u_{0}}{}^{r}{}_{\varphi} & = & \omega L \frac{L}{R} \left((\gamma-1)+\frac{r_{S}}{R}\left(1-\frac{3}{2}\gamma\right)\right), \cr
\Phi^{T}_{\omega,u_{0}}{}^{\vartheta}{}_{\varphi} & = & \omega (\gamma-1) \frac{L}{R}\sqrt{1-\left(\frac{L}{R}\right)^{2}},
\end{eqnarray}
where the notation 
$\gamma:=\frac{1}{\sqrt{1-\omega^{2}L^{2}}}$ is used. It is remarkable that 
only the components $\Phi^{T}_{\omega,u_{0}}{}^{r}{}_{\varphi}$ and 
$\Phi^{T}_{\omega,u_{0}}{}^{\varphi}{}_{r}$ depend on $r_{S}$.

In order to extract the angular velocity vector of the $u_{0}$-rotation 
generator $\Phi^{T}_{\omega,u_{0}}{}^{\mathsf{f}}{}_{\mathsf{b}}$, one needs to 
take the spatial Hodge dual in the space of $u_{0}$. This is given by the 
formula
\begin{eqnarray}
 \Omega^{T}_{\omega,u_{0}}{}^{\mathsf{f}} := \frac{1}{2}\,u_{0}{}^{\mathsf{a}}\sqrt{-\mathrm{det}(g)}\epsilon_{\mathsf{a}\mathsf{b}\mathsf{c}\mathsf{d}}\,g^{\mathsf{b}\mathsf{f}}\,\Phi^{T}_{\omega,u_{0}}{}^{\mathsf{c}}{}_{\mathsf{e}}\,g^{\mathsf{e}\mathsf{d}},
\end{eqnarray}
where $\mathrm{det}(g)$ denotes the determinant of the matrix of the metric 
$g_{\mathsf{a}\mathsf{b}}$ in our coordinates, and $\epsilon_{\mathsf{a}\mathsf{b}\mathsf{c}\mathsf{d}}$ 
is the Levi-Civita symbol.
The concrete coordinate components of the Thomas precession angular velocity vector  is
\begin{eqnarray}
 \Omega^{T}_{\omega,u_{0}}{}^{\mathsf{a}} = \left(\begin{array}{c}
 0 \cr
 \omega(\gamma-1)\sqrt{1-\frac{r_{S}}{R}}\sqrt{1-\left(\frac{L}{R}\right)^{2}} \cr
 -\omega\frac{L}{R}\frac{1}{R}\frac{1}{\sqrt{1-\frac{r_{S}}{R}}}\left((\gamma-1)+\frac{r_{S}}{R}\left(1-\frac{3}{2}\gamma\right)\right) \cr
 0 \cr
\end{array}\right).\cr
\end{eqnarray}

Let us introduce the vector fields 
$\hat{r}^{a}:=\frac{1}{\sqrt{g_{bc}\left(\partial_{r}\right)^{b}\left(\partial_{r}\right)^{c}}}\left(\partial_{r}\right)^{a}$, 
$\hat{\vartheta}^{a}:=\frac{1}{\sqrt{g_{bc}\left(\partial_{\vartheta}\right)^{b}\left(\partial_{\vartheta}\right)^{c}}}\left(\partial_{\vartheta}\right)^{a}$, 
$\hat{\varphi}^{a}:=\frac{1}{\sqrt{g_{bc}\left(\partial_{\varphi}\right)^{b}\left(\partial_{\varphi}\right)^{c}}}\left(\partial_{\varphi}\right)^{a}$, 
which are by construction an orthonormal basis in the space $E_{u_{0}}$ at 
each point, in the direction of $r$, $\vartheta$, $\varphi$. The 
metric projections of $\Omega^{T}_{\omega,u_{0}}{}^{a}$ onto this orthonormal 
basis is
\begin{eqnarray}
-g_{ab}\,\hat{r}^{a}\,\Omega^{T}_{\omega,u_{0}}{}^{b} & = & \omega(\gamma-1)\sqrt{1-\left(\frac{L}{R}\right)^{2}}, \cr
-g_{ab}\,\hat{\vartheta}^{a}\,\Omega^{T}_{\omega,u_{0}}{}^{b} & = & -\omega\frac{L}{R}\frac{1}{\sqrt{1-\frac{r_{S}}{R}}}\left((\gamma-1)+\frac{r_{S}}{R}\left(1-\frac{3}{2}\gamma\right)\right), \cr
-g_{ab}\,\hat{\varphi}^{a}\,\Omega^{T}_{\omega,u_{0}}{}^{b} & = & 0.
\label{eqOmegaTMetricProj}
\end{eqnarray}
It is remarkable that only the $\vartheta$ projection carries all the $r_{S}$ dependence.

The Thomas precession angular velocity magnitude is given by the length of the vector $\Omega^{T}_{\omega,u_{0}}{}^{\mathsf{a}}$, 
i.e., by
\begin{eqnarray}
 |\Omega|^{T}_{\omega,u_{0}} := \sqrt{-g_{\mathsf{a}\mathsf{b}}\,\Omega^{T}_{\omega,u_{0}}{}^{\mathsf{a}}\,\Omega^{T}_{\omega,u_{0}}{}^{\mathsf{b}}},
\label{eqOmega}
\end{eqnarray}
which can be evaluated to be
\begin{eqnarray}
 |\Omega|^{T}_{\omega,u_{0}} = \cr\cr
|\omega| \,\frac{1}{\sqrt{1-\frac{r_{S}}{R}}}\, 
\Bigg((\gamma-1)^{2} \cr
\qquad +\bigg(\frac{r_{S}}{R}\bigg)     \bigg(-\gamma^{2}\Big(1+2\Big(\frac{L}{R}\Big)^{2}\Big)+\gamma\Big(2+3\Big(\frac{L}{R}\Big)^{2}\Big)-\Big(1+\Big(\frac{L}{R}\Big)^{2}\Big)\bigg) \cr
\qquad +\bigg(\frac{r_{S}}{R}\bigg)^{2} \bigg(\frac{L}{R}\bigg)^{2}\bigg(\frac{3}{2}\gamma-1\bigg)^{2}
\Bigg)^{\frac{1}{2}}.
\label{eqOmegaMagn}
\end{eqnarray}

In the real experimental situation \cite{g2, mane2005} of g-2 experiments, the used 
observables are rather related to the oscillation frequency of various 
projections of the spin direction vector $w_{\omega,u_{0}}{}^{a}$, and not 
directly related to the magnitude of the precession angular velocity. Let 
$d^{a}$ be a vector field defined along the orbiting curve 
$t\mapsto\gamma_{\omega}(t)$ with the property that its coordinate components 
$d^{\mathsf{a}}(\gamma_{\omega}(t))$ are constant as a function of $t$ in our 
adapted coordinates. We call such a vector field a corotating vector. Due 
to the definition of the covariant $u_{0}$-time derivative $(\cdot)'$, it shall obey 
the equation of motion
\begin{eqnarray}
 \left(d^{\mathsf{a}}\right)^{'} & = & \Phi^{C}_{\omega,u_{0}}{}^{\mathsf{a}}{}_{\mathsf{b}}\, d^{\mathsf{b}}
\label{eqCyclicMotion}
\end{eqnarray}
with the definition 
$\Phi^{C}_{\omega,u_{0}}{}^{\mathsf{a}}{}_{\mathsf{b}}:=\frac{1}{\Lambda_{0}}\left(\dot{\gamma}_{\omega}{}^{\mathsf{c}}-\dot{\gamma}_{0}{}^{\mathsf{c}}\right)\Gamma_{\mathsf{c}\mathsf{b}}^{\mathsf{a}}$. 
That is due to Eq.(\ref{eqPrime}) and to $\frac{\mathrm{d}}{\mathrm{d}t}\,d^{\mathsf{a}}(\gamma_{\omega}(t))=0$. 
By construction or by direct substitution it is seen that the index 
pulled up version $\Phi^{C}_{\omega,u_{0}}{}^{a}{}_{b}\,g^{bc}$ is antisymmetric, 
and therefore $\Phi^{C}_{\omega,u_{0}}{}^{a}{}_{b}$ corresponds to a Lorentz 
transformation generator. Moreover $\Phi^{C}_{\omega,u_{0}}{}^{a}{}_{b}\,u_{0}{}^{b}=0$ 
holds, and therefore, it corresponds to an $u_{0}$-rotation. The equation 
of motion Eq.(\ref{eqCyclicMotion}) is therefore called the \emph{cyclic motion}. 
The coordinate components of the cyclic motion tensor are:
\begin{eqnarray}
\Phi^{C}_{\omega,u_{0}}{}^{\mathsf{a}}{}_{\mathsf{b}} & = & \left(\begin{array}{cccc}
0 & 0 & 0 & 0 \cr
0 & 0 & 0 & \Phi^{C}_{\omega,u_{0}}{}^{r}{}_{\varphi} \cr
0 & 0 & 0 & \Phi^{C}_{\omega,u_{0}}{}^{\vartheta}{}_{\varphi} \cr
0 & \Phi^{C}_{\omega,u_{0}}{}^{\varphi}{}_{r} & \Phi^{C}_{\omega,u_{0}}{}^{\varphi}{}_{\vartheta} & 0 \cr
\end{array}\right), \cr
  \mathrm{with}& & \cr
 \Phi^{C}_{\omega,u_{0}}{}^{\varphi}{}_{r} & = & -\Phi^{C}_{\omega,u_{0}}{}^{r}{}_{\varphi}\,\frac{g^{\varphi\varphi}}{g^{rr}}, \cr
 \Phi^{C}_{\omega,u_{0}}{}^{\varphi}{}_{\vartheta} & = & -\Phi^{C}_{\omega,u_{0}}{}^{\vartheta}{}_{\varphi}\,\frac{g^{\varphi\varphi}}{g^{\vartheta\vartheta}}, \cr
\Phi^{C}_{\omega,u_{0}}{}^{r}{}_{\varphi} & = & -\omega R\left(1-\frac{r_{S}}{R}\right)\left(\frac{L}{R}\right)^{2}, \cr
\Phi^{C}_{\omega,u_{0}}{}^{\vartheta}{}_{\varphi} & = & -\omega \left(\frac{L}{R}\right)\sqrt{1-\left(\frac{L}{R}\right)^{2}}.
\end{eqnarray}
This rotation generator tensor can be transformed to a more convenient form 
via an $u_{0}$ Hodge dualization
\begin{eqnarray}
 \Omega^{C}_{\omega,u_{0}}{}^{\mathsf{f}} := \frac{1}{2}\,u_{0}{}^{\mathsf{a}}\sqrt{-\mathrm{det}(g)}\epsilon_{\mathsf{a}\mathsf{b}\mathsf{c}\mathsf{d}}\,g^{\mathsf{b}\mathsf{f}}\,\Phi^{C}_{\omega,u_{0}}{}^{\mathsf{c}}{}_{\mathsf{e}}\,g^{\mathsf{e}\mathsf{d}}
\end{eqnarray}
which defines the angular velocity vector of the cyclic motion. The 
coordinate components of the cyclic angular velocity vector is
\begin{eqnarray}
 \Omega^{C}_{\omega,u_{0}}{}^{\mathsf{a}} = \left(\begin{array}{c}
 0 \cr
 -\omega \sqrt{1-\frac{r_{S}}{R}} \sqrt{1-\left(\frac{L}{R}\right)^{2}} \cr
 \omega \frac{L}{R} \frac{1}{R} \sqrt{1-\frac{r_{S}}{R}} \cr
 0 \cr
\end{array}\right).\cr
\end{eqnarray}
The observables in the real g-2 experiments are related to the oscillation 
frequency of projections of the spin direction vector $w_{\omega,u_{0}}{}^{a}$ 
onto corotating vectors, such as onto $\hat{\varphi}^{a}$. It is seen that 
\begin{eqnarray}
 \left(-g_{ab}\,d^{a}\,w_{\omega,u_{0}}{}^{b}\right)^{'} & = & -g_{ab}\,\left(\Phi^{T}_{\omega,u_{0}}-\Phi^{C}_{\omega,u_{0}}\right)^{b}{}_{c}\,d^{a}\,w_{\omega,u_{0}}{}^{c}
\end{eqnarray}
holds, where the Leibniz rule, Eq.(\ref{eqThomasMotion}), Eq.(\ref{eqCyclicMotion}), 
and the antisymmetry of $\Phi^{C}_{\omega,u_{0}}{}^{c}{}_{d}\,g^{de}$ 
has to be used. With a subsequent time derivation, one arrives at
\begin{eqnarray}
 \left(-g_{ab}\,d^{a}\,w_{\omega,u_{0}}{}^{b}\right)^{''} & = & -g_{ab}\,\left(\Phi^{T}_{\omega,u_{0}}-\Phi^{C}_{\omega,u_{0}}\right)^{b}{}_{c}\,\left(\Phi^{T}_{\omega,u_{0}}-\Phi^{C}_{\omega,u_{0}}\right)^{c}{}_{d}\,d^{a}\,w_{\omega,u_{0}}{}^{d} \cr
 & & 
\end{eqnarray}
where identity 
$\left.\left(\Phi^{T}_{\omega,u_{0}}-\Phi^{C}_{\omega,u_{0}}\right)^{a}{}_{b}\right.^{'} = \Phi^{C}_{\omega,u_{0}}{}^{a}{}_{c}\,\left(\Phi^{T}_{\omega,u_{0}}-\Phi^{C}_{\omega,u_{0}}\right)^{c}{}_{b} - \left(\Phi^{T}_{\omega,u_{0}}-\Phi^{C}_{\omega,u_{0}}\right)^{a}{}_{c}\,\Phi^{C}_{\omega,u_{0}}{}^{c}{}_{b}$ has to be used in addition. 
From now on, the corotating vector $d^{a}$ shall be assumed to be a unit 
vector, residing in $E_{u_{0}}$, and denote by $P_{(d)}{}^{a}{}_{b}:=-d^{a}g_{cb}d^{c}$ 
the orthogonal projection operator onto $d^{a}$. Since $d^{a}=P_{(d)}{}^{a}{}_{e}\,d^{e}$ 
holds, moreover since $g_{ab}\,P_{(d)}{}^{a}{}_{e}=g_{ef}\,P_{(d)}{}^{f}{}_{b}$ holds, one obtains 
the identity
\begin{eqnarray}
 \left(-g_{ab}\,d^{a}\,w_{\omega,u_{0}}{}^{b}\right)^{''} = A_{e d}\;d^{e}\,w_{\omega,u_{0}}{}^{d} \cr
 \mathrm{with} \cr
 A_{\,e d} := -g_{ef}\,\left(P_{(d)}{}^{f}{}_{b}\,\left(\Phi^{T}_{\omega,u_{0}}-\Phi^{C}_{\omega,u_{0}}\right)^{b}{}_{c}\,\left(\Phi^{T}_{\omega,u_{0}}-\Phi^{C}_{\omega,u_{0}}\right)^{c}{}_{d}\right).
\label{eqSecondDer}
\end{eqnarray}
Using now the fact that the tensor $\left(\Phi^{T}_{\omega,u_{0}}-\Phi^{C}_{\omega,u_{0}}\right)^{b}{}_{c}\,g^{cd}$ 
can be regarded as the $u_{0}$ Hodge dual of the vector $\left(\Omega^{T}_{\omega,u_{0}}-\Omega^{C}_{\omega,u_{0}}\right)^{a}$, 
the identity
\begin{eqnarray}
 A_{ed} & = & \cr
 & & -g_{ef}\,\bigg( -\left|\Omega^{T}_{\omega,u_{0}}-\Omega^{C}_{\omega,u_{0}}\right|^{2}\,P_{(d)}{}^{f}{}_{d} \cr
 & & \qquad - P_{(d)}{}^{f}{}_{b}\,\left(\Omega^{T}_{\omega,u_{0}}-\Omega^{C}_{\omega,u_{0}}\right)^{b}\,\left(\Omega^{T}_{\omega,u_{0}}-\Omega^{C}_{\omega,u_{0}}\right)^{c}\,g_{cd} \bigg)
\label{eqSecondDerTensor}
\end{eqnarray}
follows.
From Eq.(\ref{eqSecondDer}) and Eq.(\ref{eqSecondDerTensor}) it follows that
the projection $-g_{ab}\,d^{a}\,w_{\omega,u_{0}}{}^{b}$ has zero oscillation 
frequency whenever $d^{a}$ points in the direction of the vector 
$\left(\Omega^{T}_{\omega,u_{0}}-\Omega^{C}_{\omega,u_{0}}\right)^{a}$. 
Moreover, it has oscillation frequency 
$\left|\Omega^{T}_{\omega,u_{0}}-\Omega^{C}_{\omega,u_{0}}\right|$ whenever 
$d^{a}$ points in the direction $\hat{\varphi}^{a}$ (longitudinal direction), 
being orthogonal to $\left(\Omega^{T}_{\omega,u_{0}}-\Omega^{C}_{\omega,u_{0}}\right)^{a}$. 
Whenever $d^{a}$ is orthogonal to both (transverse direction), the corresponding 
projection also oscillates with $\left|\Omega^{T}_{\omega,u_{0}}-\Omega^{C}_{\omega,u_{0}}\right|$.
The experimental observables in the g-2 experiments are related to the 
oscillation frequency of the longitudinal direction, which evaluates as
\begin{eqnarray}
 \left|\Omega^{T}_{\omega,u_{0}}-\Omega^{C}_{\omega,u_{0}}\right| = 
|\omega|\, \gamma\, \sqrt{1-2\frac{r_{S}}{R}\left(\frac{L}{R}\right)^{2}\frac{1-\frac{9}{8}\frac{r_{S}}{R}}{1-\frac{r_{S}}{R}}}.
\label{eqTLongOsc}
\end{eqnarray}
It is interesting to note that from the metric projections of the vector 
$\left(\Omega^{T}_{\omega,u_{0}}-\Omega^{C}_{\omega,u_{0}}\right)^{a}$, only 
the $\vartheta$ projection has dependence on $r_{S}$:
\begin{eqnarray}
-g_{ab}\,\hat{r}^{a}\,(\Omega^{T}_{\omega,u_{0}}-\Omega^{C}_{\omega,u_{0}}){}^{b} & = & \omega\gamma\sqrt{1-\left(\frac{L}{R}\right)^{2}}, \cr
-g_{ab}\,\hat{\vartheta}^{a}\,(\Omega^{T}_{\omega,u_{0}}-\Omega^{C}_{\omega,u_{0}}){}^{b} & = & -\omega\gamma\frac{L}{R}\left(1-\frac{3}{2}\frac{r_{S}}{R}\right)\frac{1}{\sqrt{1-\frac{r_{S}}{R}}}, \cr
-g_{ab}\,\hat{\varphi}^{a}\,(\Omega^{T}_{\omega,u_{0}}-\Omega^{C}_{\omega,u_{0}}){}^{b} & = & 0.
\label{eqTLongProjs}
\end{eqnarray}

All these expressions were derived and cross-checked using the GRTensorII Maple package 
\cite{grtensor}. The above calculations will also be made available as 
supplementary material.

\section{Evaluation of the GR correction for Thomas precession}
\label{secEvaluationT}

As shown in the previous section, the expression for the precession angular velocity 
$|\Omega|^{T}_{\omega,u_{0}}$ can be obtained as an analytical formula Eq.(\ref{eqOmegaMagn}). Its 
Minkowski limit is 
\begin{eqnarray}
 |\Omega|^{T}_{\omega,u_{0}}\Big\vert_{r_{S}=0} & = & \;|\omega|\,(\gamma-1) \;=\; |\omega|\,\frac{\beta^{2}\,\gamma^{2}}{1+\gamma},
\label{eqMink}
\end{eqnarray}
with the notation $\beta:=\omega\,L$. 
This is the special relativistic formula for Thomas precession, presented 
also in many textbooks \cite{rindler2006, matolcsi2007}.
The first order correction of GR can be obtained via taking the first 
Taylor term of $|\Omega|^{T}_{\omega,u_{0}}$ as a function of $r_{S}$. 
The first order absolute error turns out to be:
\begin{eqnarray}
 r_{S}\,\left(\left.\frac{\mathrm{d}}{\mathrm{d}r_{S}}|\Omega|^{T}_{\omega,u_{0}}\right\vert_{r_{S}=0}\right) = 
 \,-  \frac{ r_{S}}{2R}\, \frac{L^2}{R^2}\, |\omega|\,\frac{2\gamma^2+\gamma-1}{1+\gamma}.
\label{eqErr}
\end{eqnarray}
The quotient of Eq.(\ref{eqErr}) and Eq.(\ref{eqMink}) gives the relative systematic 
error of $|\Omega|^{T}_{\omega,u_{0}}$ due to neglection of GR, which evaluates to
\begin{eqnarray}
 \,-  \frac{ r_{S}}{2R}\, \frac{L^2}{R^2}\, \frac{2\gamma^2+\gamma-1}{\beta^{2}\gamma^{2}}.
\end{eqnarray}
It is seen that in the ultrarelativistic limit ($|\beta|\rightarrow 1$), it 
evaluates to $-\frac{ r_{S}}{R}\,\frac{L^2}{R^2}$, whereas in the nonrelativistic 
limit ($|\beta|\rightarrow 0$) it can be approximated as 
$-\frac{ r_{S}}{R}\,\frac{L^2}{R^2}\,\frac{1}{\beta^{2}}$. It is quite 
remarkable that the relative systematic error at the nonrelativistic limit 
diverges as $\frac{1}{\beta^{2}}$, which can be understood by recalling that 
the small special relativistic phenomenon of Thomas precession is competing 
against the small GR correction in that regime, and the GR correction happens 
to win in that limit. An other feature of the GR correction is that the vector
$\Omega^{T}_{\omega,u_{0}}{}^{a}$ tilts with respect to the Minkowski limit. 
To the first order, the tilt angle can be estimated from the first Taylor 
term of Eq.(\ref{eqOmegaTMetricProj}) in terms of $r_{S}$ and from Eq.(\ref{eqMink}), 
and is seen to be:
\begin{eqnarray}
 \frac{r_{S}}{R}\frac{L}{R}\frac{\left(\gamma-\frac{1}{2}\right)(\gamma+1)}{\beta^{2}\gamma^{2}}.
\end{eqnarray}
It is seen that in the ultrarelativistic limit the tilt angle becomes 
$\frac{r_{S}}{R}\frac{L}{R}$, whereas in the nonrelativistic limit it becomes 
$\frac{r_{S}}{R}\frac{L}{R}\frac{1}{\beta^{2}}$. When the oscillation frequency 
Eq.(\ref{eqTLongOsc}) of the longitudinal or transverse spin projection is considered, it behaves as
\begin{eqnarray}
 \left|\Omega^{T}_{\omega,u_{0}}-\Omega^{C}_{\omega,u_{0}}\right| \Big\vert_{r_{S}=0} = |\omega|\,\gamma
\label{eqTMink}
\end{eqnarray}
in the Minkowski limit, and its first order correction is
\begin{eqnarray}
 r_{S}\,\left(\left.\frac{\mathrm{d}}{\mathrm{d}r_{S}}\left|\Omega^{T}_{\omega,u_{0}}-\Omega^{C}_{\omega,u_{0}}\right| \right\vert_{r_{S}=0}\right) = -\frac{r_{S}}{R}\left(\frac{L}{R}\right)^{2}|\omega|\gamma,
\end{eqnarray}
and therefore its relative systematic error is $-\frac{r_{S}}{R}\left(\frac{L}{R}\right)^{2}$, 
independent of the velocity $|\beta|$. The vector $\left(\Omega^{T}_{\omega,u_{0}}-\Omega^{C}_{\omega,u_{0}}\right)^{a}$ 
also suffers a tilt, and the tilt angle is seen to be 
\begin{eqnarray}
 -\frac{r_{S}}{R}\frac{L}{R}
\end{eqnarray}
from Eq.(\ref{eqTMink}) and from Eq.(\ref{eqTLongProjs}), independently of the 
velocity $|\beta|$.

In the real muon g-2 experiment, the muons are injected with a relativistic Lorentz dilatation factor 
$\gamma\approx29.3$ to the storage ring \cite{g2, mane2005}. This means a velocity relative 
to the speed of light $|\beta|\approx0.999417412329374$, i.e., the muons are 
ultrarelativistic and one is in the $|\beta|\rightarrow1$ limit. Therefore, 
the Thomas precession frequency as well as the longitudinal spin oscillation 
frequency is modified by a relative systematic error of
\begin{eqnarray}
 - \frac{r_{S}}{R}\frac{L^2}{R^2}
\label{eqRelErrUltraRel}
\end{eqnarray}
in the muon g-2 experiment. 
Using now Eq.(\ref{eqRelErrUltraRel}) and the radius and Schwarzschild radius of the Earth, $R\approx 6.371\cdot10^{6}\,\mathrm{m}$ 
and $r_{S}\approx 9\cdot10^{-3}\,\mathrm{m}$, and the storage ring radius $L\approx 7.5\,\mathrm{m}$, one infers 
that the relative systematic error made when neglecting GR in the estimation of Thomas precession 
frequency is
\begin{eqnarray}
 \approx -2 \cdot10^{-21},
\end{eqnarray}
which is negligible, given the present experimental accuracy. The most sensitive 
observable to the GR correction seems to be the tilt of the spin precession 
plane against the nominal horizontal plane, which is 
${\approx}1.7\cdot 10^{-15}\,\frac{1}{\beta^{2}}$ radians with the parameters of the existing 
g-2 magnet setup. It is seen that an experimental detection of GR effects are 
only possible with sufficiently low $|\beta|$ and sufficiently large angular 
resolution, which is not realistic e.g.\ in a muon g-2 experiment. 
Moreover in the above corrections the effects of GR on the electrodynamics 
of the beam optics is not yet accounted for, which we quantify in the coming sections.

\section{Electromagnetic fields in an idealized storage ring over Schwarzschild}
\label{secEldin}

In this section, the electromagnetic fields in an idealized storage ring 
\cite{mane2005} over a Schwarzschild background spacetime is outlined, which guides the 
particle beam along the nominal beam trajectory Eq.(\ref{eqWorldline}).

\begin{figure}[!h]
\begin{center}
\includegraphics[width=3.6cm]{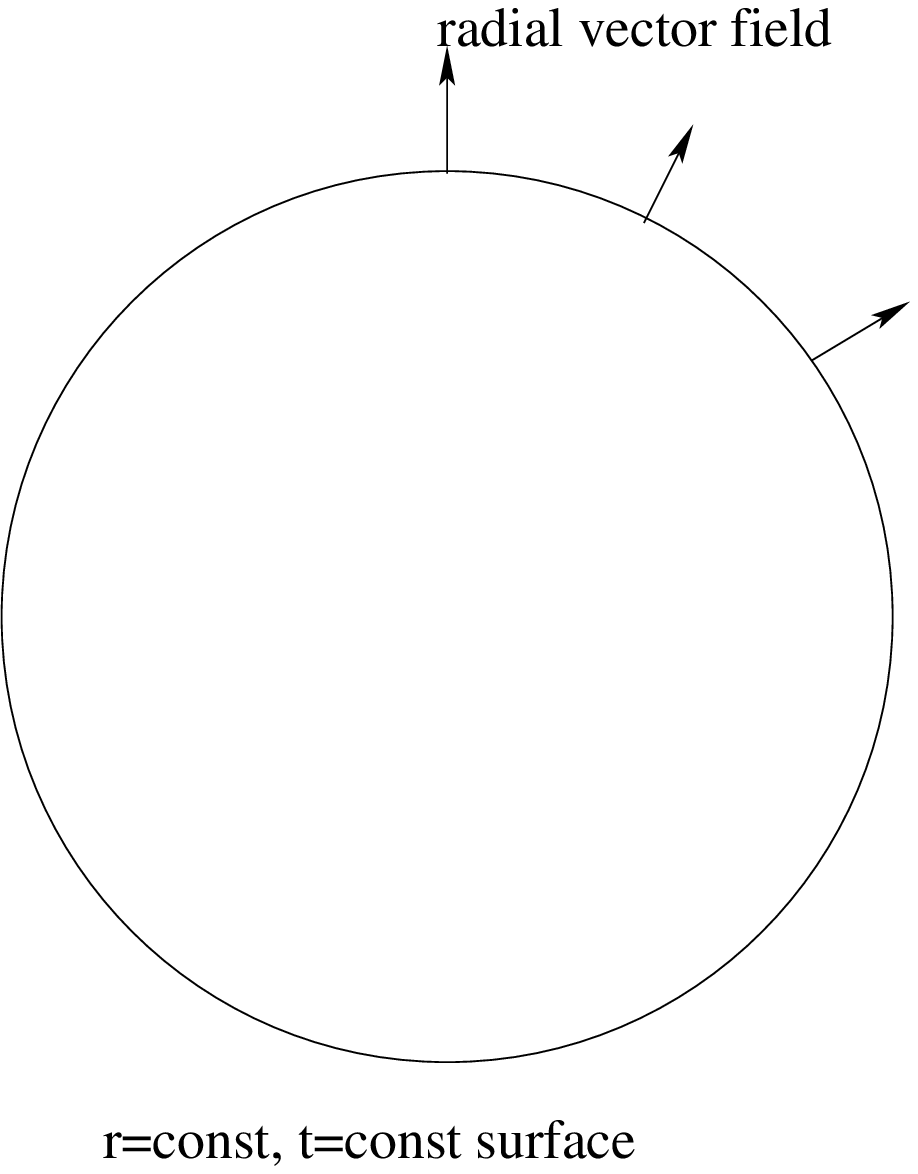}\hspace*{2cm}\includegraphics[width=5cm]{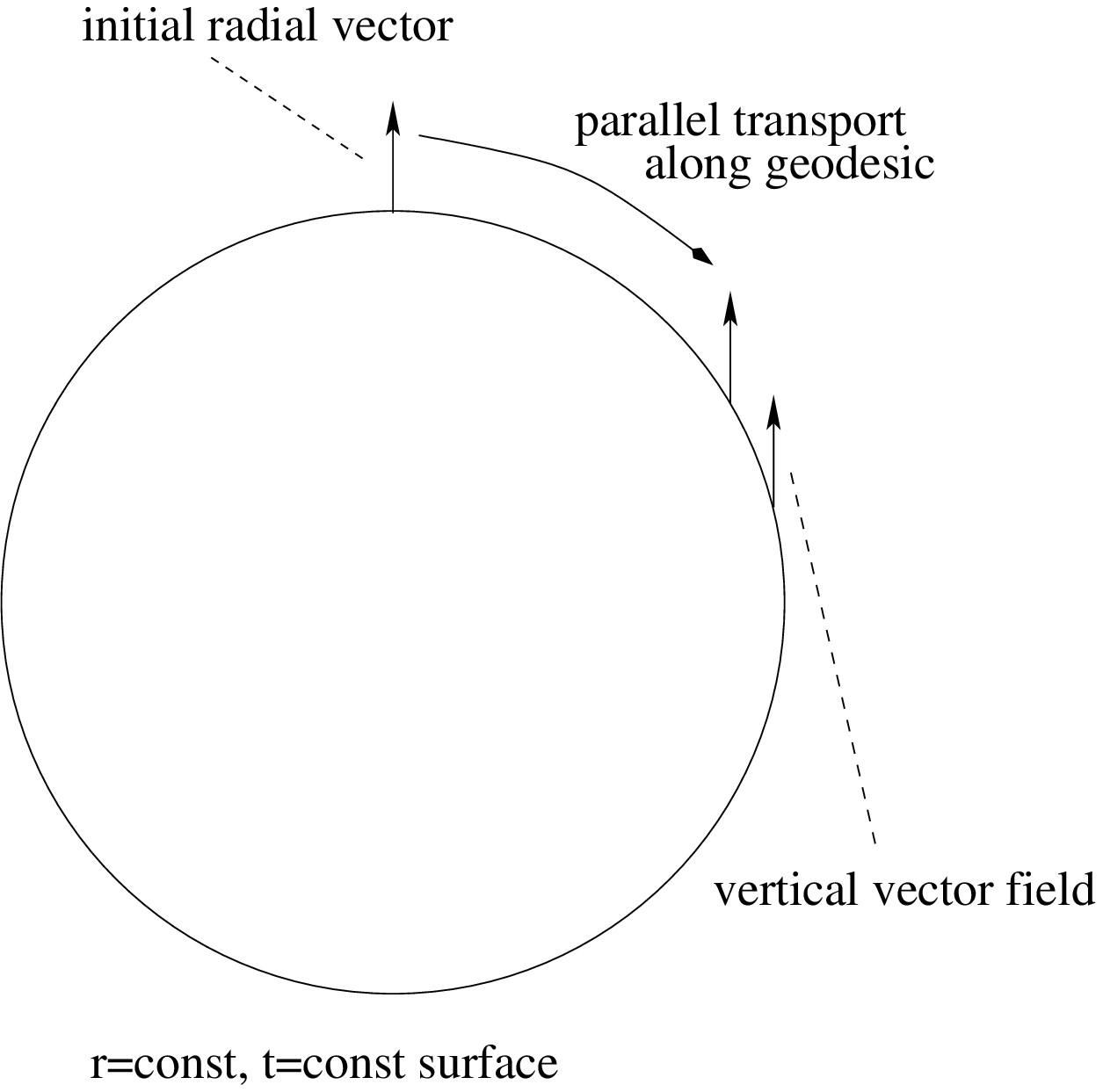}
\end{center}
\caption{Left panel: illustration of the outward pointing radial unit vector 
field $\hat{r}^{a}$ on an $r=\mathrm{const}$, $t=\mathrm{const}$ surface. 
Right panel: illustration of the geometric construction of the vertical unit vector field 
$\hat{v}^{a}$ on an $r=\mathrm{const}$, $t=\mathrm{const}$ surface. The outward 
pointing radial unit vector field is taken at a preferred point, and it is 
parallel transported along geodesics to other points of the surface, in terms 
of the embedded flat metric.}
\label{figVertical}
\end{figure}

The most important electromagnetic field in a storage ring is usually, a homogeneous 
magnetic field, which forces the particles onto a closed circular orbit at 
the nominal trajectory. The first task is to define such a field configuration 
over our Schwarzschild background. In order to do that, recall that the 
outward pointing unit normal vector field of the surface $r=\mathrm{const}$ 
surface is called the radial unit vector field, and is denoted by $\hat{r}^{a}$. 
Its components in our coordinate conventions are
\begin{eqnarray}
 \hat{r}^{\mathsf{a}}(t,r,\vartheta,\varphi) = \left(\begin{array}{c} 0 \cr \sqrt{1-\frac{r_{S}}{r}} \cr 0 \cr 0 \cr \end{array}\right).
\end{eqnarray}
The intersection of an $r=\mathrm{const}$ surface with a Killing time $t=\mathrm{const}$ 
surface is a two-sphere. The induced metric $g_{ab}+\hat{r}_{a}\hat{r}_{b}$ on it is just the round two-sphere 
metric with radius $r$. Besides the induced metric, on the pertinent 
surface the embedded flat metric can also be defined by 
$g_{ab}+\hat{r}_{a}\hat{r}_{b}-\mathrm{d}r_{a}\mathrm{d}r_{b}$, where $\mathrm{d}r$ 
denotes the exterior derivative of the radius function $r$. 
This surface has a vector field on it defined by taking the radial unit vector field at a preferred point, and parallel transporting it to any point 
of the two-sphere along a geodesic, in terms of the embedded flat metric. 
That vector field, when initialized from the storage ring axis, i.e., from $\vartheta=0$, is denoted by 
$\hat{v}^{a}$, and is called the vertical unit vector field. Its components 
in our coordinate conventions are
\begin{eqnarray}
 \hat{v}^{\mathsf{a}}(t,r,\vartheta,\varphi) = \left(\begin{array}{c} 0 \cr \cos\vartheta \cr -\frac{1}{r}\sin\vartheta \cr 0 \cr \end{array}\right).
\end{eqnarray}
Note that the unit length and the parallel transport is understood in the 
embedded flat metric on the surface, and not in the merely restricted 
Schwarzschild metric to the surface. These vector fields and their 
geometric construction is illustrated in Fig.~\ref{figVertical}.

The static, axisymmetric, homogeneous vertical magnetic vector field $B^{a}$, 
which forces the particles on a cyclotronic motion in the storage ring, 
is simply defined by a radial scalar multiple of the vertical unit vector 
field $\hat{v}^{a}$. Namely, one has the ansatz 
$B^{a}(t,r,\vartheta,\varphi)=b(r)\,\hat{v}^{a}(t,r,\vartheta,\varphi)$ with 
$r\mapsto b(r)$ being a radial function such that the electromagnetic field 
corresponding to $B^{a}$ satisfies the vacuum Maxwell equations \cite{hanni1976}. 
This implies that in our coordinates it has the components 
\begin{eqnarray}
 B^{\mathsf{a}}(t,r,\vartheta,\varphi) = B\,\sqrt{\frac{1-\frac{r_{S}}{r}}{1-\frac{r_{S}}{R} \left(\frac{L}{R}\right)^{2}}}\, \left(\begin{array}{c} 0 \cr \cos\vartheta \cr -\frac{1}{r}\,\sin\vartheta \cr 0 \end{array}\right),
\end{eqnarray}
with $B$ being a constant. It is normalized such that on the 
$r=R$ and $\vartheta=\Theta$ surface, it has pseudolength equal to the 
constant $|B|$. The 
magnetic component of our electromagnetic field is defined by the Hodge dual 
of $B^{\mathsf{a}}$ in the laboratory frame $u_{0}$:
\begin{eqnarray}
 F^{B}_{\mathsf{b}\mathsf{c}} := -u_{0}{}^{\mathsf{a}}\,\sqrt{-\mathrm{det}(g)}\epsilon_{\mathsf{a}\mathsf{b}\mathsf{c}\mathsf{d}}\,B^{\mathsf{d}}.
\end{eqnarray}
Direct substitution shows that $F^{B}_{bc}$ satisfies the vacuum Maxwell equations 
over Schwarzschild spacetime.

In the muon g-2 or EDM experimental settings \cite{mane2005}, an electrostatic beam focusing 
optics is used to maintain the beam stability. Since this beam optics is 
standing on the Earth's surface together with the storage ring, on the average 
it exerts a radial electrostatic field at the nominal trajectory, balancing the beam against 
falling towards the Earth. This electrostatic field, being radial, has the form 
$E_{R}{}^{a}(t,r,\vartheta,\varphi)=e(r)\,\hat{r}^{a}(t,r,\vartheta,\varphi)$, where 
the radial function $r\mapsto e(r)$ has to be chosen such that the electromagnetic 
field corresponding to $E_{R}{}^{a}$ satisfies the vacuum Maxwell equations. This 
implies that in our coordinates it has components
\begin{eqnarray}
 E_{R}{}^{\mathsf{a}}(t,r,\vartheta,\varphi) = E_{R}\,\frac{R^{2}}{r^{2}}\, \left(\begin{array}{c} 0 \cr \sqrt{1-\frac{r_{S}}{r}} \cr 0 \cr 0 \end{array}\right),
\end{eqnarray}
with $E_{R}$ being a constant. It is normalized such that on the 
$r=R$ surface, it has pseudolength equal to the constant $\left|E_{R}\right|$. The 
electric component of our electromagnetic field is defined by 
$E_{R}{}^{\mathsf{a}}$ and $u_{0}$:
\begin{eqnarray}
 F^{E_{R}}_{ab} := u_{0}{}^{c}g_{ca}E_{R}{}^{d}g_{db}-u_{0}{}^{c}g_{cb}E_{R}{}^{d}g_{da}.
\end{eqnarray}
Direct substitution shows that $F^{E_{R}}_{ab}$ satisfies the vacuum Maxwell equations over 
Schwarzschild spacetime.

\begin{figure}[!h]
\begin{center}
\includegraphics[width=8.0cm]{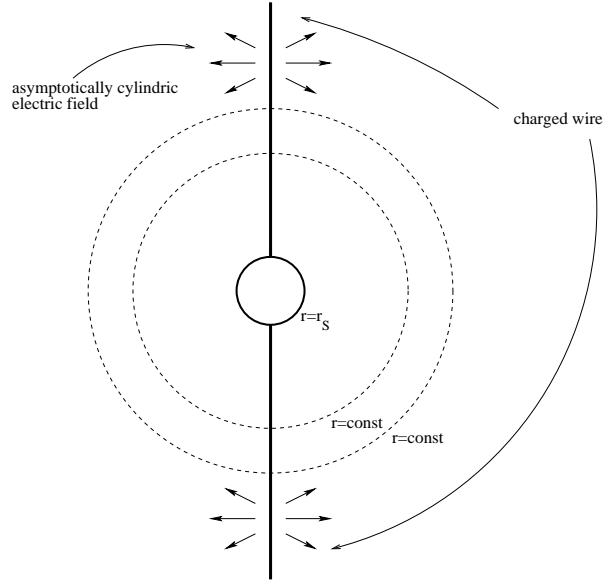}
\end{center}
\caption{Illustration of the asymptotically cylindric electrostatic field of a uniformly charged 
wire suspended in a Schwarzschild spacetime. This field models the asymptotically horizontal 
electric field in an electrostatic or in a mixed magnetic-electric storage ring. 
The charge density of the wire can be determined by the charge integral under 
$r=\mathrm{const}$ two-spheres, as described in the text. The free 
integration constants of such Maxwell fields are set so that the wire has uniform charge density in 
terms of $r$, and that under the sphere $r=r_{S}$ one has vanishing total charge.}
\label{figCylindricE}
\end{figure}

In the EDM experimental settings \cite{mane2005}, sometimes a mixed electrostatic-magnetostatic 
storage ring is used, or a pure electrostatic storage ring. These employ an 
outward pointing cylindrical electrostatic field, being the electric field 
of a uniformly charged infinite wire in an idealized model. The field of such a wire 
could be directly calculated, since the Green's function of electrostatics/magnetostatics 
is well understood over a Schwarzschild background \cite{binet1976}. The evaluation of 
the pertinent Green's integral over the wire, however, becomes quite complicated 
if one would like an analytical formula. Therefore, it is easier to 
construct the pertinent field configuration using symmetry ansatzes. As it is well-known, 
in the Minkowski limit, the field of such a uniformly charged wire, comoving with the observer $u_{0}$, 
and placed at $\vartheta=0$ and $\vartheta=\pi$ is known to be a cylindric one, with field strength decaying 
like ${\sim}\frac{1}{r\,\sin\vartheta}$, and its electrostatic potential 
being of the form ${\sim}\ln\left(r\,\sin\vartheta\right)+\mathrm{const}$. 
Our ansatz over the Schwarzschild background therefore shall be that the 
electromagnetic vector potential has the form
\begin{eqnarray}
 A_{\mathsf{a}}(t,r,\vartheta,\varphi) & = & \left(\begin{array}{cccc} f(r)\,\ln(\sin\vartheta)+g(r), & 0, & 0, & 0 \end{array}\right)
\end{eqnarray}
which should solve the vacuum Maxwell equations for $r>r_{S}$, outside the 
wire singularity $\vartheta=0$ and $\vartheta=\pi$. That requirement determines the unknown 
radial functions $f(r)$ and $g(r)$ up to four integration constants $a,b,c,d$, 
and one arrives at the solution
\begin{eqnarray}
 A_{t}(t,r,\vartheta,\varphi) &=& a\,\left(\ln(\sin\vartheta)+\left(1-\frac{r_{S}}{r}\right)\ln(r-r_{S})\right) \cr
 & &   \quad +b\,\left(\frac{1}{r}\ln(\sin\vartheta)-\frac{1}{r_{S}}\ln(r)+\frac{1}{r_{S}}\left(1-\frac{r_{S}}{r}\right)\ln(r-r_{S})\right) \cr
 & &   \quad\quad -c\,\frac{1}{r}+d.
\label{eqSolA}
\end{eqnarray}
Since the field strength tensor is the exterior derivative of $A_{a}$, it is 
invariant to the choice of the integration constant $d$, and thus one may set 
$d=0$ by convention. Also, the Maxwell's equations are linear, and thus the 
shape of the field configuration is invariant to the normalization of the 
solution, and therefore, e.g., one may set $a=1$ by convention, in order to 
determine the shape of the field. Given these choices, the total 
charge within an $r=\mathrm{const}$ sphere of a solution Eq.(\ref{eqSolA}) 
evaluates as $\int_{S^{2}_{(r=\mathrm{const)}}} (*\mathrm{d}A) = 
2\pi\,\left(2\left(-r-(b+r_{S})\,\ln(r-r_{S})-c\right)+b\,2\,(\ln(2)-1)\right)$. 
If one requires that the charge under the $r=\mathrm{const}$ two-spheres 
should not diverge as $r\rightarrow r_{S}$, the equality $b=-r_{S}$ follows. 
With this condition, the charge under an $r=\mathrm{const}$ two-sphere 
shall be $2\pi\,\left(2\left(-r-c\right)-r_{S}\,2\,(\ln(2)-1)\right)$, meaning 
that the charge density in terms of $r$ along the wire is uniform, 
as illustrated in Fig.~\ref{figCylindricE}. The last 
integration constant $c$ can be fixed by requiring that the charge under 
the sphere $r=r_{S}$ must vanish, and then $c=-r_{S}\,\ln(2)$ follows. 
Thus, the field strength tensor $F^{E_{H}}_{ab}$ of a uniformly charged 
vertical wire suspended over a Schwarzschild background becomes a constant factor times 
$\mathrm{d}A$, with $A$ taken from 
Eq.(\ref{eqSolA}) and $a=1$, $b=-r_{S}$, $c=-r_{S}\,\ln(2)$, $d=0$. If one wishes to express that in terms of an electric vector 
field $E_{H}{}^{a}$ in the space of the observer $u_{0}$, defined in terms of $F^{E_{H}}_{ab}=u_{0}{}^{c}g_{ca}E_{H}{}^{d}g_{db}-u_{0}{}^{c}g_{cb}E_{H}{}^{d}g_{da}$, 
then one arrives at the expression
\begin{eqnarray}
 E_{H}{}^{\mathsf{a}}(t,r,\vartheta,\varphi) = E_{H} \,\frac{L}{r\,\sin\vartheta}\,\sqrt{1-\frac{r_{S}}{r}}\,\N_{r_{S}}\, \left(\begin{array}{c} 0 \cr \sin\vartheta\,\left(1+\frac{r_{S}}{r}\,\ln(\frac{1}{2}\sin\vartheta)\right) \cr \frac{1}{r}\,\cos\vartheta \cr 0 \end{array}\right), \cr\cr
\qquad\mathrm{with}\; \N_{r_{S}} := \Big(\Big(\frac{L}{R}\Big)^{2}\Big(1+\frac{r_{S}}{R}\ln\Big(\frac{L}{2R}\Big)\Big)^{2}+\Big(1-\Big(\frac{L}{R}\Big)^{2}\Big)\Big(1-\frac{r_{S}}{R}\Big)\Big)^{-\frac{1}{2}}\cr
\end{eqnarray}
where the normalization factor $E_{H}$ is a real constant. The symbol 
$\N_{r_{S}}$ was introduced for brevity. The formula is normalized such 
that at the $r=R$, $\vartheta=\Theta$ location of the nominal beam line, the 
vector field $E_{H}{}^{a}$ has pseudolength equal to the constant $\left|E_{H}\right|$. As a cross-check, 
direct evaluation also shows that the field strength tensor 
$F^{E_{H}}_{ab}=u_{0}{}^{c}g_{ca}E_{H}{}^{d}g_{db}-u_{0}{}^{c}g_{cb}E_{H}{}^{d}g_{da}$ 
indeed satisfies the vacuum Maxwell equations over Schwarzschild spacetime.

The total electromagnetic field strength tensor is then 
$F_{ab}:=F^{B}_{ab}+F^{E_{H}}_{ab}+F^{E_{R}}_{ab}$. In our coordinates, it has the components:
\newline\hspace*{-16mm}\begin{minipage}{\textwidth}
\begin{eqnarray}
 F_{\mathsf{a}\mathsf{b}}(t,r,\vartheta,\varphi) = \cr\cr
\qquad\left(\begin{array}{cccc} 
 0 & -E_{R}\,\frac{R^{2}}{r^{2}} & 0 & 0 \cr
 E_{R}\,\frac{R^{2}}{r^{2}} & 0 & 0 & -B\,\frac{r\,\sin^{2}\vartheta}{\sqrt{1-\frac{r_{S}}{R}\left(\frac{L}{R}\right)^{2}}} \cr
 0 & 0 & 0 & -B\,\frac{r^{2}\,\sin\vartheta\,\cos\vartheta}{\sqrt{1-\frac{r_{S}}{R}\left(\frac{L}{R}\right)^{2}}} \cr
 0 & B\,\frac{r\,\sin^{2}\vartheta}{\sqrt{1-\frac{r_{S}}{R}\left(\frac{L}{R}\right)^{2}}} & B\,\frac{r^{2}\,\sin\vartheta\,\cos\vartheta}{\sqrt{1-\frac{r_{S}}{R}\left(\frac{L}{R}\right)^{2}}} & 0 \cr
\end{array}\right) \cr\cr\cr
+E_{H}\,L\,\N_{r_{S}}\,\left(\begin{array}{cccc} 
 0 & -\frac{1}{r}\big(1+\frac{r_{S}}{r}\ln\big(\frac{1}{2}\sin\vartheta\big)\big) & -\big(1-\frac{r_{S}}{r}\big)\,\frac{\cos\vartheta}{\sin\vartheta} & 0 \cr
 \frac{1}{r}\big(1+\frac{r_{S}}{r}\ln\big(\frac{1}{2}\sin\vartheta\big)\big) & 0 & 0 & 0 \cr
 \big(1-\frac{r_{S}}{r}\big)\,\frac{\cos\vartheta}{\sin\vartheta} & 0 & 0 & 0 \cr
 0 & 0 & 0 & 0 \cr
\end{array}\right). \cr
\label{eqF}
\end{eqnarray}
\end{minipage}\newline
If the storage ring is purely magnetic, then $B\neq0$, $E_{H}=0$, $E_{R}\neq0$, 
whereas for a purely electrostatic storage ring one has $B=0$, $E_{H}\neq0$, $E_{R}\neq0$, 
while for a mixed magnetic-electric ring one has $B\neq0$, $E_{H}\neq0$, $E_{R}\neq0$.

\section{The absolute Larmor rotation in presence of electromagnetic fields}
\label{secAbsLarmor}

Whenever electromagnetic field is also present, the BMT equation, i.e., the 
second equation in Eq.(\ref{eqNewtonBMT}), causes further rotation of the spin 
direction vector in addition to the Thomas rotation, the so-called \emph{Larmor rotation}.

The beam motion is modelled by the cyclotronic worldline 
Eq.(\ref{eqWorldline}), and it has to satisfy the Newton equation, i.e., 
the first line of Eq.(\ref{eqNewtonBMT}), against the field strength tensor 
Eq.(\ref{eqF}). Given $\omega$ and $E_{H}$, that is fulfilled whenever the consistency conditions
\begin{eqnarray}
 B & = & \omega\,\frac{m\,\gamma}{q}\,\sqrt{1-\frac{r_{S}}{R}\left(\frac{L}{R}\right)^{2}} \;+\; \frac{E_{H}}{\omega\,L}\,\N_{r_{S}}\,\sqrt{1-\frac{r_{S}}{R}}\,\sqrt{1-\frac{r_{S}}{R}\left(\frac{L}{R}\right)^{2}}, \cr
 E_{R} & = & \frac{r_{S}}{R}\,\frac{m\,\gamma}{2\,q\,R}\,\frac{1}{\sqrt{1-\frac{r_{S}}{R}}} \;-\; E_{H}\,\frac{L}{R}\,\frac{r_{S}}{R}\,\N_{r_{S}}\,\left(1+\ln\left(\frac{L}{2R}\right)\right)
\label{eqCyclotron}
\end{eqnarray}
hold with the notation $\gamma:=\frac{1}{\sqrt{1-\omega^{2}L^{2}}}$. 
These are the equations of cyclotronic motion over a Schwarzschild background 
spacetime, with $\omega$ being the cyclotron circular frequency in terms of 
$u_{0}$-proper time. From this point on, it is assumed that Eq.(\ref{eqCyclotron}) 
is satisfied, i.e., that the motion Eq.(\ref{eqWorldline}) of the particle 
is a consequence of a cyclotronic motion in electromagnetic field over a Schwarzschild spacetime. 
In the real g-2 and EDM experiments \cite{g2, mane2005} the magnetic field strength $B$ 
and the horizontal electric field strength $E_{H}$ are the fixed (measured) 
parameters, which determine $\omega$ and $E_{R}$, given the constants 
$\frac{m}{q}$, $L$, $R$, and the GR correction parameter $\frac{r_{S}}{R}$.

We are now at the point of evaluating the Larmor tensor 
$L_{\omega}{}^{b}{}_{d}:=-\Lambda_{\omega}\,\frac{\mu}{s}\,\left(g^{bc}F_{cd}-u_{\omega}{}^{b}u_{\omega}{}^{c}F_{cd}-g^{bc}F_{ce}u_{\omega}{}^{e}u_{\omega}{}^{f}g_{fd}\right)$ 
which gives a contribution to the BMT equation, i.e., to the second line of Eq.(\ref{eqNewtonBMT}). 
With that definition, the spin transport equation reads as
\begin{eqnarray}
 \frac{\mathrm{d}}{\mathrm{d}t} \tilde{w}_{\omega}{}^{\mathsf{b}}(t) = \F_{\omega}{}^{\mathsf{b}}{}_{\mathsf{c}}\,\tilde{w}_{\omega}{}^{\mathsf{c}}(t) \;+\; L_{\omega}{}^{\mathsf{b}}{}_{\mathsf{c}}\,\tilde{w}_{\omega}{}^{\mathsf{c}}(t)
\label{eqAbsLarmor}
\end{eqnarray}
in terms of Killing time. In our coordinate conventions it has components as

{
\hspace*{-2.8cm}\begin{minipage}{\textwidth}
\begin{eqnarray}
L_{\omega}{}^{\mathsf{b}}{}_{\mathsf{d}}=-\frac{1}{2}\frac{\g\,\omega\,\sqrt{1-\frac{r_{S}}{R}}}{1-\omega^{2}L^{2}} \cr\cr\cr
\quad\left(\begin{array}{cccc}
0 & -\omega L \frac{L}{R}\frac{\left(1-\frac{3}{2}\frac{r_{S}}{R}\right)}{\left(1-\frac{r_{S}}{R}\right)^{\frac{3}{2}}} & -\omega L R \frac{\sqrt{1-\left(\frac{L}{R}\right)^{2}}}{\left(1-\frac{r_{S}}{R}\right)^{\frac{1}{2}}} & 0 \cr
-\omega L \frac{L}{R} \left(1-\frac{3}{2}\frac{r_{S}}{R}\right)\left(1-\frac{r_{S}}{R}\right)^{\frac{1}{2}} & 0 & 0 & L\frac{L}{R}\left(1-\frac{3}{2}\frac{r_{S}}{R}\right) \cr
 & & & \cr
-\omega \frac{L}{R}\sqrt{1-\left(\frac{L}{R}\right)^{2}}\left(1-\frac{r_{S}}{R}\right)^{\frac{1}{2}} & 0 & 0 & \frac{L}{R}\sqrt{1-\left(\frac{L}{R}\right)^{2}} \cr
0 & -\frac{1}{R}\frac{\left(1-\frac{3}{2}\frac{r_{S}}{R}\right)}{\left(1-\frac{r_{S}}{R}\right)} & -\frac{R}{L}\sqrt{1-\left(\frac{L}{R}\right)^{2}} & 0 \cr
\end{array}\right)\cr\cr\cr
- \frac{\g\,q\,E_{H}\,\sqrt{1-\omega^{2}L^{2}}}{2\,m\,\omega\,L}\,\N_{r_{S}}\,\left(1-\frac{r_{S}}{R}\right)^{2} \cr
\quad\left(\begin{array}{cccc}
0 & -\omega L \frac{L}{R}\frac{1}{\left(1-\frac{r_{S}}{R}\right)^{\frac{3}{2}}} & -\omega L R \frac{\sqrt{1-\left(\frac{L}{R}\right)^{2}}}{\left(1-\frac{r_{S}}{R}\right)^{\frac{3}{2}}} & 0 \cr
-\omega L \frac{L}{R}\left(1-\frac{r_{S}}{R}\right)^{\frac{1}{2}} & 0 & 0 & L\frac{L}{R} \cr
 & & & \cr
-\omega \frac{L}{R} \frac{\sqrt{1-\left(\frac{L}{R}\right)^{2}}}{\left(1-\frac{r_{S}}{R}\right)^{\frac{1}{2}}} & 0 & 0 & \frac{L}{R}\frac{\sqrt{1-\left(\frac{L}{R}\right)^{2}}}{1-\frac{r_{S}}{R}} \cr
0 & -\frac{1}{R}\frac{1}{\left(1-\frac{r_{S}}{R}\right)} & -\frac{R}{L}\frac{\sqrt{1-\left(\frac{L}{R}\right)^{2}}}{1-\frac{r_{S}}{R}} & 0 \cr
\end{array}\right)\cr\cr 
\end{eqnarray}
\end{minipage}
}

\noindent
where the usual definition of the gyromagnetic factor $\g:=\frac{2\,m\,\mu}{q\,s}$ 
was used. By construction, or by direct substitution it is seen that the 
tensor $L_{\omega}{}^{a}{}_{b}\,g^{bc}$ is antisymmetric, and therefore 
$L_{\omega}{}^{a}{}_{b}$ corresponds to a Lorentz transformation generator. 
Moreover, $L_{\omega}{}^{a}{}_{b}\,u_{\omega}{}^{b}=0$ holds, which means that 
$L_{\omega}{}^{a}{}_{b}$ describes an $u_{\omega}{}^{a}$-rotation generator. 
That phenomenon is the \emph{Larmor rotation}, and is an absolute, 
observer independent effect.

\section{The relative Larmor precession as seen by the laboratory observer}
\label{secRelLarmor}

Because of Eq.(\ref{eqAbsLarmor}), according to the laboratory observer $u_{0}$, the spin direction vector 
$w_{\omega,u_{0}}{}^{a}$ satisfies the equation 
\begin{eqnarray}
 \left(w_{\omega,u_{0}}{}^{f}\right)' & = & \Phi^{T}_{\omega,u_{0}}{}^{f}{}_{b}\,w_{\omega,u_{0}}{}^{b} \;+\; \Phi^{L}_{\omega,u_{0}}{}^{f}{}_{b}\,w_{\omega,u_{0}}{}^{b},
\end{eqnarray}
with the definition 
$\Phi^{L}_{\omega,u_{0}}{}^{f}{}_{b}:=\frac{1}{\Lambda_{0}}\,B_{u_{0},u_{\omega}}{}^{f}{}_{c}\,L_{\omega}{}^{c}{}_{a}\,B_{u_{\omega},u_{0}}{}^{a}{}_{b}$. 
By construction or by direct substitution it is seen that 
$\Phi^{L}_{\omega,u_{0}}{}^{f}{}_{b}\,g^{bc}$ is antisymmetric, which means 
that it describes a Lorentz transformation generator. Moreover, 
$\Phi^{L}_{\omega,u_{0}}{}^{f}{}_{b}\,u_{0}{}^{b}=0$ holds, which means 
that it describes an $u_{0}$-rotation generator. 
The effect of $\Phi^{L}_{\omega,u_{0}}{}^{f}{}_{b}$ is 
called the \emph{Larmor precession}. 
In our coordinate conventions, it has the components
\begin{eqnarray}
\Phi^{L}_{\omega,u_{0}}{}^{\mathsf{a}}{}_{\mathsf{b}} & = & \left(\begin{array}{cccc}
0 & 0 & 0 & 0 \cr
0 & 0 & 0 & \Phi^{L}_{\omega,u_{0}}{}^{r}{}_{\varphi} \cr
0 & 0 & 0 & \Phi^{L}_{\omega,u_{0}}{}^{\vartheta}{}_{\varphi} \cr
0 & \Phi^{L}_{\omega,u_{0}}{}^{\varphi}{}_{r} & \Phi^{L}_{\omega,u_{0}}{}^{\varphi}{}_{\vartheta} & 0 \cr
\end{array}\right), \cr
  \mathrm{with}& & \cr
 \Phi^{L}_{\omega,u_{0}}{}^{\varphi}{}_{r} & = & -\Phi^{L}_{\omega,u_{0}}{}^{r}{}_{\varphi}\,\frac{g^{\varphi\varphi}}{g^{rr}}, \cr
 \Phi^{L}_{\omega,u_{0}}{}^{\varphi}{}_{\vartheta} & = & -\Phi^{L}_{\omega,u_{0}}{}^{\vartheta}{}_{\varphi}\,\frac{g^{\varphi\varphi}}{g^{\vartheta\vartheta}}, \cr
\Phi^{L}_{\omega,u_{0}}{}^{r}{}_{\varphi} & = & -\frac{1}{2}\,\g\,\omega\,L\,\gamma\,\frac{L}{R}\,\left(1-\frac{3}{2}\frac{r_{S}}{R}\right) \;-\; \frac{\g\,q}{2\,m}\,\frac{E_{H}\,L\,\N_{r_{S}}\,\left(1-\frac{r_{S}}{R}\right)^{\frac{3}{2}}}{\omega\,R\,\gamma^{2}} ,\cr
\Phi^{L}_{\omega,u_{0}}{}^{\vartheta}{}_{\varphi} & = & -\frac{1}{2}\,\g\,\omega\,\gamma\,\frac{L}{R}\,\sqrt{1-\frac{L^2}{R^{2}}} \;-\; \frac{\g\,q}{2\,m}\,\frac{E_{H}\,\N_{r_{S}}\,\sqrt{1-\frac{r_{S}}{R}}}{\omega\,R\,\gamma^{2}}\,\sqrt{1-\frac{L^{2}}{R^{2}}}
\end{eqnarray}
with using the notation $\gamma:=\frac{1}{\sqrt{1-\omega^{2}L^{2}}}$.

The angular velocity vector of the Larmor precession can be obtained by the 
$u_{0}$ spatial Hodge dual of $\Phi^{L}_{\omega,u_{0}}{}^{a}{}_{b}$, according 
to the definition
\begin{eqnarray}
 \Omega^{L}_{\omega,u_{0}}{}^{\mathsf{f}} := \frac{1}{2}\,u_{0}{}^{\mathsf{a}}\sqrt{-\mathrm{det}(g)}\epsilon_{\mathsf{a}\mathsf{b}\mathsf{c}\mathsf{d}}\,g^{\mathsf{b}\mathsf{f}}\,\Phi^{L}_{\omega,u_{0}}{}^{\mathsf{c}}{}_{\mathsf{e}}\,g^{\mathsf{e}\mathsf{d}}.
\end{eqnarray}
Its components in our coordinate conventions are
\begin{eqnarray}
 \Omega^{L}_{\omega,u_{0}}{}^{\mathsf{a}} = \left(\begin{array}{c}
 0 \cr
 -\frac{1}{2}\,\g\,\omega\,\gamma\,\sqrt{1-\frac{L^{2}}{R^{2}}}\sqrt{1-\frac{r_{S}}{R}} \;-\; \frac{\g\,q}{2\,m}\,\frac{E_{H}\,\N_{r_{S}}\,\left(1-\frac{r_{S}}{R}\right)\sqrt{1-\frac{L^{2}}{R^{2}}}}{\omega\,L\,\gamma^{2}} \cr
 \frac{1}{2}\,\g\,\omega\,\gamma\,\frac{1}{R}\,\frac{L}{R}\,\frac{1-\frac{3}{2}\frac{r_{S}}{R}}{\sqrt{1-\frac{r_{S}}{R}}} \;+\; \frac{\g\,q}{2\,m}\,\frac{E_{H}\,\N_{r_{S}}\,\left(1-\frac{r_{S}}{R}\right)}{\omega\,R^{2}\,\gamma^{2}} \cr
 0 \cr
\end{array}\right).\cr
\end{eqnarray}

For charged particles in a cyclotronic motion, the total angular velocity vector of the spin precession is determined by 
$\Omega^{S}_{\omega,u_{0}}{}^{a}:=\Omega^{T}_{\omega,u_{0}}{}^{a}+\Omega^{L}_{\omega,u_{0}}{}^{a}$. 
Its components in our coordinate conventions are
\newline\hspace*{-11mm}\begin{minipage}{\textwidth}
\begin{eqnarray}
 \Omega^{S}_{\omega,u_{0}}{}^{\mathsf{a}} = \left(\begin{array}{c}
 0 \cr
 -\omega\,(1+\gamma\,a)\,\sqrt{1-\frac{r_{S}}{R}}\,\sqrt{1-\frac{L^2}{R^2}} \;-\; (1+a)\,\frac{E_{H}\,q\,\N_{r_{S}}\,\left(1-\frac{r_{S}}{R}\right)\sqrt{1-\frac{L^{2}}{R^{2}}}}{m\,\omega\,L\,\gamma^{2}} \cr
 \omega\,\frac{1}{R}\,\frac{L}{R}\,\left(\sqrt{1-\frac{r_{S}}{R}}+\gamma\,a\,\frac{1-\frac{3}{2}\frac{r_{S}}{R}}{\sqrt{1-\frac{r_{S}}{R}}}\right) \;+\; (1+a)\frac{E_{H}\,q\,\N_{r_{S}}\,\left(1-\frac{r_{S}}{R}\right)}{m\,\omega\,R^{2}\,\gamma^{2}} \cr
 0 \cr
\end{array}\right), \cr 
\end{eqnarray}
\end{minipage}\newline
where the usual notation $a:=\frac{\g-2}{2}$ was used for the magnetic moment 
anomaly. The metric projections onto the orthonormal frame $\hat{r}^{a}$, 
$\hat{\vartheta}^{a}$, $\hat{\varphi}^{a}$ of $\Omega^{S}_{\omega,u_{0}}{}^{a}$ 
are the followings:
\newline\hspace*{-15mm}\begin{minipage}{\textwidth}
\begin{eqnarray}
 -g_{ab}\,\hat{r}^{a}\,\Omega^{S}_{\omega,u_{0}}{}^{b} & = & -\omega\,(1+\gamma\,a)\,\sqrt{1-\frac{L^2}{R^2}} \;-\; (1+a)\,\frac{E_{H}\,q\,\N_{r_{S}}\,\left(1-\frac{r_{S}}{R}\right)^{\frac{1}{2}}\sqrt{1-\frac{L^{2}}{R^{2}}}}{m\,\omega\,L\,\gamma^{2}} , \cr
 -g_{ab}\,\hat{\vartheta}^{a}\,\Omega^{S}_{\omega,u_{0}}{}^{b} & = & \omega\,\frac{L}{R}\,\left(\sqrt{1-\frac{r_{S}}{R}}+\gamma\,a\,\frac{1-\frac{3}{2}\frac{r_{S}}{R}}{\sqrt{1-\frac{r_{S}}{R}}}\right) \;+\; (1+a)\frac{E_{H}\,q\,\N_{r_{S}}\,\left(1-\frac{r_{S}}{R}\right)}{m\,\omega\,R\,\gamma^{2}} , \cr
 -g_{ab}\,\hat{\varphi}^{a}\,\Omega^{S}_{\omega,u_{0}}{}^{b} & = & 0.
\label{eqOmegaSProjections}
\end{eqnarray}
\end{minipage}\newline

As discussed at the end of Section~\ref{secRelFWtransport}, the oscillation frequencies of the longitudinal and 
transverse spin component is determined by the length of the vector 
$\left(\Omega^{S}_{\omega,u_{0}}-\Omega^{C}_{\omega,u_{0}}\right)^{a}$. 
The coordinate components of that vector is given by
\begin{eqnarray}
 \left(\Omega^{S}_{\omega,u_{0}}-\Omega^{C}_{\omega,u_{0}}\right)^{\mathsf{a}} = 
\omega\,\gamma\,a\, \left(\begin{array}{c}
 0 \cr
 -\left(1-\frac{r_{S}}{R}\right)^{\frac{1}{2}}\,\sqrt{1-\frac{L^2}{R^2}} \cr
 \frac{1}{R}\,\frac{L}{R}\,\frac{1-\frac{3}{2}\frac{r_{S}}{R}}{\left(1-\frac{r_{S}}{R}\right)^{\frac{1}{2}}} \cr
 0 \cr
\end{array}\right)\cr
\qquad\qquad\qquad + (1+a)\,\frac{E_{H}\,q}{m\,\omega\,L\,\gamma^{2}} \,\left(\begin{array}{c} 
 0 \cr
 -\N_{r_{S}}\,\left(1-\frac{r_{S}}{R}\right)\,\sqrt{1-\frac{L^{2}}{R^{2}}} \cr
 \frac{1}{R}\,\frac{L}{R}\,\N_{r_{S}}\,\left(1-\frac{r_{S}}{R}\right) \cr
 0 \cr
\end{array}\right).
\label{eqOmegaSMinusOmegaC}
\end{eqnarray}
Its metric projections onto $\hat{r}^{a}$, $\hat{\vartheta}^{a}$, $\hat{\varphi}^{a}$ is
\begin{eqnarray}
 -g_{ab}\,\hat{r}^{a}\,\left(\Omega^{S}_{\omega,u_{0}}-\Omega^{C}_{\omega,u_{0}}\right)^{b} & = & -\omega\,\gamma\,a\,\sqrt{1-\frac{L^2}{R^2}} \cr
 & & \;\;\;-\; (1+a)\,\frac{E_{H}\,q\,\N_{r_{S}}\,\left(1-\frac{r_{S}}{R}\right)^{\frac{1}{2}}}{m\,\omega\,L\,\gamma^{2}}\,\sqrt{1-\frac{L^{2}}{R^{2}}} , \cr
 -g_{ab}\,\hat{\vartheta}^{a}\,\left(\Omega^{S}_{\omega,u_{0}}-\Omega^{C}_{\omega,u_{0}}\right)^{b} & = & \omega\,\gamma\,a\,\frac{L}{R}\,\frac{1-\frac{3}{2}\frac{r_{S}}{R}}{\left(1-\frac{r_{S}}{R}\right)^{\frac{1}{2}}} \cr
 & & \;\;\;+\; (1+a)\,\frac{E_{H}\,q\,\N_{r_{S}}\,\left(1-\frac{r_{S}}{R}\right)}{m\,\omega\,L\,\gamma^{2}}\,\frac{L}{R} , \cr
 -g_{ab}\,\hat{\varphi}^{a}\,\left(\Omega^{S}_{\omega,u_{0}}-\Omega^{C}_{\omega,u_{0}}\right)^{b} & = & 0.
\end{eqnarray}

In the g-2 experiments \cite{g2, mane2005}, one has $B\neq 0$ and $E_{H}=0$, i.e., purely 
magnetic storage ring is used, and the main observable is
\begin{eqnarray}
 \left\vert\Omega^{S}_{\omega,u_{0}}-\Omega^{C}_{\omega,u_{0}}\right\vert = |\omega\,a|\,\gamma\,\Bigg(
\bigg(1-\frac{L^{2}}{R^{2}}\bigg) + \frac{L^{2}}{R^{2}}\frac{\big(1-\frac{3}{2}\frac{r_{S}}{R}\big)^{2}}{1-\frac{r_{S}}{R}}
\Bigg)^{\frac{1}{2}}.
\label{eqg2freq}
\end{eqnarray}
It is seen, that GR gives slight contribution, the quantification of which 
is done in the following section.

In electric dipole moment (EDM) search experiments 
\cite{senichev2017, semertzidis2016, talman2017}, the so called \emph{frozen spin method} 
is used: the parameters $B$ and $E_{H}$ are adjusted such, that the vector 
$\left(\Omega^{S}_{\omega,u_{0}}-\Omega^{C}_{\omega,u_{0}}\right)^{a}$ 
vanishes to a best possible accuracy, and thus the observed spin direction 
vector is always exactly tangential to the orbit, i.e., points in the direction 
$\hat{\varphi}^{a}$. If an EDM of the particle would exist, it manifests as a 
residual precession, out of the orbital plane. 
The identities Eq.(\ref{eqCyclotron}) and Eq.(\ref{eqOmegaSMinusOmegaC}) 
tell us that in the Minkowski limit ($r_{S}=0$), the frozen spin condition is quite 
possible to achieve via setting
$E_{H}=-\frac{a}{1+a}\,\frac{m}{q}\,L\,\omega^{2}\,\gamma^{3}$, assuming 
that the magnetic anomaly $a$ of the particle was measured in advance. 
In the GR case, however, the complete vanishing of 
$\left(\Omega^{S}_{\omega,u_{0}}-\Omega^{C}_{\omega,u_{0}}\right)^{a}$ is not 
possible to achieve exactly, and at best the frozen spin condition can only 
be achieved approximately. For instance, one may require that the vertical 
projection $-g_{ab}\,\hat{v}^{a}\,\left(\Omega^{S}_{\omega,u_{0}}-\Omega^{C}_{\omega,u_{0}}\right)^{b}$ 
vanishes, which is fulfilled whenever
\begin{eqnarray}
 E_{H} & = & -\frac{a}{1+a}\,\frac{m}{q}\,L\,\omega^{2}\,\gamma^{3}\,\frac{1}{\N_{r_{S}}\,\left(1-\frac{r_{S}}{R}\right)^{\frac{1}{2}}}\,\frac{1-\frac{3}{2}\frac{r_{S}}{R}\frac{L^{2}}{R^{2}}}{1-\frac{r_{S}}{R}\frac{L^{2}}{R^{2}}}
\label{eqFrozenSpinCondition}
\end{eqnarray}
holds, as seen from Eq.(\ref{eqOmegaSMinusOmegaC}). In that case, there will be a residual precession out of the orbital 
plane, with magnitude
\begin{eqnarray}
 \left\vert\Omega^{S}_{\omega,u_{0}}-\Omega^{C}_{\omega,u_{0}}\right\vert & = & \vert\omega\,a\vert\,\gamma\, \frac{1}{2}\,\frac{r_{S}}{R}\,\frac{L}{R}\, \left(1-\frac{r_{S}}{R}\right)^{-\frac{1}{2}}\,\frac{\sqrt{1-\frac{L^{2}}{R^{2}}}}{\sqrt{1-\frac{r_{S}}{R}\frac{L^{2}}{R^{2}}}}
\label{eqEDMGR}
\end{eqnarray}
and the direction of the precession is upward vertical whenever 
$-\omega\,a$ is positive, and downward vertical otherwise. 
The magnitude of this GR correction to a frozen spin scenario is quantified in the next section.

\section{Evaluation of the GR corrections to Thomas plus Larmor precession}
\label{secEvaluationLarmor}

First, the effect of GR is evaluated on the cyclotron frequency $\omega$. 
This frequency is uniquely determined by the first line of Eq.(\ref{eqCyclotron}), 
given the vertical magnetic field strength $B$ and the horizontal electric field 
strength $E_{H}$ of the storage ring, and the constants 
$\frac{m}{q}$, $L$, $R$, along with the GR correction parameter $\frac{r_{S}}{R}$. 
In order to determine the $r_{S}$ dependence of $\omega$, it should be regarded 
as an implicit function of $r_{S}$. Its first order GR correction is nothing 
but its first Taylor expansion term in terms of $r_{S}$, and that can be easily 
derived by differentiating against $r_{S}$ the pertinent consistency equation 
between $\omega$, $B$ and $E_{H}$. It follows that one has
\begin{eqnarray}
 r_{S}\,\left(\left.\frac{\mathrm{d}}{\mathrm{d}r_{S}}\,\omega \right\vert_{r_{S}=0} \right) & = & \omega\,\frac{1}{2}\,\frac{r_{S}}{R}\,\frac{L^{2}}{R^{2}}\,\frac{1-\frac{1}{\gamma^{2}}+\frac{2\,E_{H}\,q\,L\,\left(1+\ln\left(\frac{L}{2R}\right)\right)}{m\,\gamma}}{\gamma^{2}-1-\frac{E_{H}\,q\,L}{m\,\gamma}}.
\label{eqomegacorr}
\end{eqnarray}
The last quotient expression is ${\approx}1$ for relativistic particles, and thus 
the relative error of $\omega$ caused by the neglection of GR is 
${\approx}\frac{1}{2}\,\frac{r_{S}}{R}\,\frac{L^{2}}{R^{2}}$, which is of the 
order of $10^{-21}$, as already quantified in Section~\ref{secEvaluationT}. 
Therefore, that correction is pretty much negligible for current and foreseeable 
experimental applications. 
One could also say, that in the system there are two small parameters: 
$\frac{r_{S}}{R}$ and $\frac{L}{R}$, and the corrections of the third order, 
such as $\frac{r_{S}}{R}\,\frac{L^{2}}{R^{2}}$, are considered to be negligible. 
That also means that in the GR correction estimations, the $r_{S}$ dependence 
of $\omega$ can be neglected in order to simplify the formulas.

As a next step, the effect of GR on the main observable of the muon g-2 experiments \cite{g2, mane2005} is 
evaluated, which is the frequency of the longitudinal spin projection. 
That was shown equal to the quantity 
$\left\vert\Omega^{S}_{\omega,u_{0}}-\Omega^{C}_{\omega,u_{0}}\right\vert$, 
which was shown to take the form Eq.(\ref{eqg2freq}) in a purely magnetic g-2 ring ($B\neq 0$, $E_{H}=0$). 
In the Minkowski limit, this has its well known special relativistic form
\begin{eqnarray}
 \left\vert\Omega^{S}_{\omega,u_{0}}-\Omega^{C}_{\omega,u_{0}}\right\vert\,\Big\vert_{r_{S}=0} & = & \vert \omega\,a\vert\,\gamma,
\end{eqnarray}
which is the principal formula for g-2 determination. Its first order Taylor 
expansion term in terms of $r_{S}$ is seen to be
\begin{eqnarray}
 r_{S}\,\left(\left.\frac{\mathrm{d}}{\mathrm{d}r_{S}}\left|\Omega^{S}_{\omega,u_{0}}-\Omega^{C}_{\omega,u_{0}}\right| \,\right\vert_{r_{S}=0}\right) = -|\omega\,a|\,\gamma\,\frac{r_{S}}{R}\,\frac{L^{2}}{R^{2}},
\end{eqnarray}
causing a relative systematic error of $-\frac{r_{S}}{R}\,\frac{L^{2}}{R^{2}}$, 
being negligible, of the order of $-2\cdot 10^{-21}$. 
Therefore, that correction is negligible for current g-2 experimental applications. 
(For simplicity, the tiny $r_{S}$ dependent correction of $\omega$, quantified in Eq.(\ref{eqomegacorr}), was neglected in the formula.)

The first apparent sizable GR correction can be seen in the total 
spin precession vector $\Omega^{S}_{\omega,u_{0}}{}^{a}$. Namely, whenever 
one uses a purely magnetic storage ring ($B\neq 0$, $E_{H}=0$) with a particle with 
magnetic anomaly $-1<a<0$, such as deuteron nuclei which do have 
$a\approx-0.142$ \cite{stone2005}, then there exists a unique velocity (or $B$) 
setting for which $1+a\,\gamma=0$ holds. As seen from Eq.(\ref{eqOmegaSProjections}), in that 
case the spin precession vector $\Omega^{S}_{\omega,u_{0}}{}^{a}$ vanishes 
in the Minkowski limit ($r_{S}=0$), meaning that the spin direction vector 
is merely parallel transported in the space of the laboratory observer. 
One could call such a scenario a ``\emph{zero torque setting}''. 
In the GR case, however, a small approximately 
horizontal component of $\Omega^{S}_{\omega,u_{0}}{}^{a}$ persists, of the magnitude 
$\left\vert\Omega^{S}_{\omega,u_{0}}\right\vert=\left\vert\omega\,\frac{L}{R}\,\left(\sqrt{1-\frac{r_{S}}{R}}-\frac{1-\frac{3}{2}\frac{r_{S}}{R}}{\sqrt{1-\frac{r_{S}}{R}}}\right)\right\vert$, 
as seen from Eq.(\ref{eqOmegaSProjections}).
Taking its first Taylor term in $r_{S}$, this is of the magnitude $\beta\,\frac{r_{S}}{2R^{2}}$ with 
the notation $\beta:=\omega\,L$, being the circular velocity in units of 
speed of light. When translated to ordinary units from the used geometric units, 
this is nothing but $\beta\,\frac{\mathbf{g}}{c}$ with $\mathbf{g}$ denoting 
the gravitational acceleration on the Earth's surface, and $c$ denoting the 
speed of light. This residual precession quantifies as 
${\approx}3.26\cdot 10^{-8}\,\mathrm{rad/sec}$ if one assumes a deuteron 
beam, which shall satisfy $1+a\,\gamma=0$ at $\beta\approx 0.9804$. In such a 
configuration with an initially tangential spin vector, the spin will not merely be parallelly transported as in the 
Minkowski limit, but it will tilt in and out of the orbital plane. The 
amplitude of this vertical polarization buildup is, however, too small to experimentally 
measure. That is because the spin vector is not approximately perpendicular to the 
residual $\Omega^{S}_{\omega,u_{0}}{}^{a}$ vector at all times, and thus the polarization 
does not accumulate with elapsing time. That issue can be remedied with an 
experimental setting discussed below.

\begin{figure}[!h]
\begin{center}
\includegraphics[width=9cm]{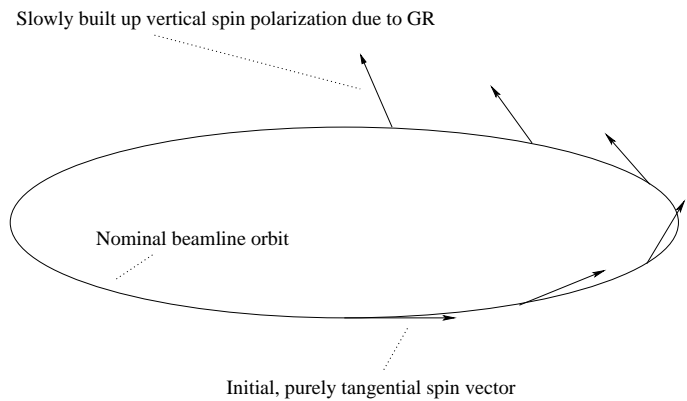}
\end{center}
\caption{Illustration of the effect of GR corrections to a frozen spin 
setting. The GR correction results in a slow buildup of vertical spin 
polarization, if the beam was initially longitudinally polarized. The rate 
of the polarization buildup is given by Eq.(\ref{eqGReffect}).}
\label{figGReffect}
\end{figure}

The GR correction is expected to have a measurable effect on \emph{frozen spin experiments}, such as EDM 
search experiments \cite{senichev2017, semertzidis2016, talman2017}. For these settings, 
in the Minkowski limit, the spin direction vector always follows the tangent 
of the orbit curve, i.e., 
$\left(\Omega^{S}_{\omega,u_{0}}-\Omega^{C}_{\omega,u_{0}}\right)^{a}\Big\vert_{r_{S}=0}=0$ holds. 
In case of GR, there are no such parameter settings when this frozen spin 
condition holds, but as indicated previously, one may can choose a setting 
in which case the vertical projection of the vector 
$\left(\Omega^{S}_{\omega,u_{0}}-\Omega^{C}_{\omega,u_{0}}\right)^{a}$ vanishes. 
In that case, the residual precession shall elevate the spin direction vector 
out of the orbital plane, with a rate Eq.(\ref{eqEDMGR}). Again, in order to 
quantify that, we take its first Taylor expansion term in terms of $r_{S}$, 
which is
\begin{eqnarray}
 r_{S}\,\left(\left.\frac{\mathrm{d}}{\mathrm{d}r_{S}}\left|\Omega^{S}_{\omega,u_{0}}-\Omega^{C}_{\omega,u_{0}}\right| \,\right\vert_{r_{S}=0}\right) & = & \vert\omega\,a\vert\,\gamma\, \frac{1}{2}\,\frac{r_{S}}{R}\,\frac{L}{R}\,\sqrt{1-\frac{L^{2}}{R^{2}}}.
\end{eqnarray}
Taking into account that $\frac{L}{R}\ll 1$ and $\beta=\omega\,L$, this quantifies as $\vert a\,\beta\gamma\vert\,\frac{r_{S}}{2R^{2}}$. 
When translated from the geometric units to normal units, it corresponds to the 
expression
\begin{eqnarray}
 \vert a\,\beta\gamma\vert\,\frac{\mathbf{g}}{c}
\label{eqGReffect}
\end{eqnarray}
which quantifies as $\approx\vert a\,\beta\gamma\vert\,3.26\cdot 10^{-8}\,\mathrm{rad/sec}$.\footnote{Our 
formula Eq.(\ref{eqGReffect}) is also in accordance with the 
post-Newtonian approximative results of \cite{orlov2012}, which was obtained 
for the special case of a purely electrostatic ($B=0$, $E_{H}\neq 0$) frozen 
spin storage ring. In such a ring, the Newton equation 
Eq.(\ref{eqCyclotron}) and the frozen spin condition 
Eq.(\ref{eqFrozenSpinCondition}) as well as the zero magnetic field condition 
$B=0$ needs to be satisfied at the same time, in the $r_{S}=0$ limit. That 
is only possible for particles with $a>0$, and also singles out a 
so-called \emph{magic momentum} $\beta\,\gamma=\frac{1}{\sqrt{a}}$ which 
is necessary in order to satisfy all these conditions. The residual vertical precession 
due to GR contribution is seen in that special case to be $\sqrt{a}\,\frac{\mathbf{g}}{c}$, 
by means of Eq.(\ref{eqGReffect}), which confirms the calculation of \cite{orlov2012}. 
It is seen, however, that using a mixed magnetic-electric frozen spin storage 
ring might be more advantageous, since it can be applied to any $a$, and 
does not single out a small value for the Lorentz factor $\gamma$. Indeed, 
via using large $\gamma$, the GR effect can be arbitrarily magnified. 
Therefore, for a specific GR experiment, we propose to use a mixed 
magnetic-electric frozen spin ring, with as large $\gamma$ as possible, and particles with large $a$.} 
Thus, in this setting, given an initially longitudinally polarized beam, a 
vertical polarization buildup will be seen at the rate of 
$\left\vert\Omega^{S}_{\omega,u_{0}}-\Omega^{C}_{\omega,u_{0}}\right\vert\approx\vert a\,\beta\gamma\vert\,32.6\,\mathrm{nrad/sec}$, 
as shown in Fig.~\ref{figGReffect}. 
This magnitude of polarization buildup rate is well within the experimental 
reach for the planned EDM experiments 
\cite{senichev2017, semertzidis2016, talman2017}, since the spin coherence 
of e.g.\ proton beams can be maintained for about an hour with 
today's beam technology. Therefore, we propose to use the foreseen \emph{frozen spin} 
or EDM experiment also as sensitive GR experiments on spin propagation. 
Although the fake EDM signal by GR is much larger than an expected true EDM 
signal, they are distinguishable due to their opposite sign flip 
behavior for a change of the beam circulation direction, i.e., due to their 
opposite chirality behavior.

\section{Concluding remarks}
\label{secConclusion}

In this paper, a fully general relativistic calculation was performed in order 
to evaluate the GR corrections over a Schwarzschild background for particle 
spin precession experiments, such as the muon g-2 measurement \cite{g2} or 
the electric dipole moment (EDM) search experiments \cite{senichev2017, semertzidis2016, talman2017}. 
It turns out that although GR gives a first order correction in $r_S$ for certain important observables, its contribution 
to the muon g-2 measurements is negligible. That is because in that setting 
spin precession in the horizontal plane is studied, and GR mainly gives contribution 
to the other direction, moreover the amount of 
precession due to the g-2 signal is large anyway. However, for EDM search experiments, in which 
case a so-called frozen spin situation is obtained with a combination of the 
$\gamma$ factor and of the electric and magnetic fields, the GR effects imitate 
a fake EDM signal. That is a precession out of the orbital plane, of the magnitude 
$\vert a\,\beta\gamma\vert\,\mathbf{g}/c$, being of the order of $3.26\cdot 10^{-8}\,\mathrm{rad/sec}$, which 
is well above the planned sensitivity of ${\approx}10^{-9}\,\mathrm{rad/sec}$ of 
these experiments, and pretty much above the Standard Model (SM) expectation. Thus, the 
EDM experiments provide a unique platform for experimental test of GR in terms 
of particle spin propagation. Moreover, in any frozen spin experiment with 
aimed precision not worse than $10^{-6}\,\mathrm{rad/sec}$, the 
GR signal needs to be subtracted in order to access any signature coming from 
particle physics --- SM or BSM. Fortunately, direct experimental elimination of such 
backgrounds may also be achieved by comparing the effect as seen from normally 
and reversely rotating beams, since the GR effect 
would change sign, but the EDM effect would not \cite{senichev2017, semertzidis2016, talman2017}. If such subtraction is 
done, the GR signal is expected to be distinguishable from a real EDM signal 
and vice versa.

\section*{Acknowledgments}

The authors would like to thank Ferenc Sikl\'er and Dezs\H{o} Horv\'ath for 
the motivating discussions. The very useful technical discussions with 
Richard M.\ Talman on the frozen spin EDM experiments and with 
Zolt\'an Keresztes on the BMT equation is also greatly acknowledged. 
This work was supported in part by the Hungarian Scientific Research fund 
(NKFIH 123959) and the János Bolyai Research Scholarship of the Hungarian 
Academy of Sciences.

\end{document}